\documentclass[a4paper,11pt]{article}
\usepackage[utf8]{inputenc}
\usepackage{color}
\usepackage{cite}
\usepackage{hyperref}
\usepackage{amsmath}
\usepackage{graphicx}

\usepackage[top=0.4in, bottom=0.7in, left=0.60in, right=0.5in]{geometry}

\begin{document}

\title{{\bf\Large{$E1$ and M1 radiative transitions involving heavy-light axial, pseudoscalar and vector quarkonia  in the framework of Bethe-Salpeter equation}}}
\author{Vaishali Guleria$^1$, Eshete Gebrehana$^2$, Shashank Bhatnagar$^1$}

\maketitle \small{$^1$Department of Physics, University Institute of Sciences, Chandigarh University, Mohali-140413, India\\
$^2$ Department of Physics, Woldia University, Woldia, Ethiopia\\}

\begin{abstract}
\normalsize{This work is an extension of our previous work in \cite{bhatnagar20} to calculate M1 transitions, $0^{-+}\rightarrow 1^{--} \gamma$, and E1 transitions involving axial vector mesons such as, $1^{+-} \rightarrow 0^{-+}\gamma$, and $0^{-+}\rightarrow 1^{+-} \gamma $ for which very little data is available as of now. We make use of the general structure of the transition amplitude, $M_{fi}$ derived in our previous work \cite{bhatnagar20} as a linear superposition of terms involving all possible combinations of $++$, and $--$ components of Salpeter wave functions of final and initial hadrons. In the present work, we make use of leading Dirac structures in the hadronic Bethe-Salpeter wave functions of the involved hadrons, which makes the formulation more rigorous. We evaluate the decay widths for both the above mentioned $M1$ and $E1$ transitions. We have used algebraic forms of Salpeter wave functions obtained through analytic solutions of mass spectral equations for ground and excited states of $1^{--}$,$0^{-+}$ and $1^{+-}$ heavy-light quarkonia in approximate harmonic oscillator basis to do analytic calculations of their decay widths. We have compared our results with experimental data, where ever available, and other models.}
\end{abstract}

\section{Introduction}
One of the challenging areas in hadronic physics is  probing the inner structure of hadrons. Charmonium occupies an intermediate regime between the $b\bar{b}$ system and the light mesons. Radiative decays of charmonia are good testing grounds for various models, due to the fact that emitted photon can be directly detected, and electromagnetic interactions are well understood.

Radiative transitions characterized by $\Delta L=0$ are the magnetic dipole, M1 transitions, while those characterized by $|\Delta L|=1$ are the electric dipole, E1 transitions. The M1 transition mode is sensitive to relativistic effects, specially between different spatial multiplets (where $n > n'$), while the E1 transitions are much stronger than M1 transitions, and involve transitions between excited states. In this work, besides the M1 transitions, we study the E1 transitions involving $1^{+-}$ mesons, which are the P-wave states.

The P wave $c\bar{c}$ states were ﬁrst observed in 1976 by the SLAC-LBL experiment at SLAC/SPEAR \cite{slaca,slacb}, where they observed the decay, $\Psi(3684) \rightarrow \gamma + \chi_c$. The P wave $b\bar{b}$ states were ﬁrst observed by the Columbia-Stony Brook (CUSB) experiment at the Cornell CESR electron-positron storage ring \cite{cusb1,cusb2} and conﬁrmed by the CLEO experiment at CESR \cite{cleoa}.

An indirect way of producing P- wave states is through $e^- e^+$ annihilation, which produces
$^3S_1$ $(1^{--})$ charmonium states such as,  $J/\Psi(1S)$ and $\Psi(2S)$. Then the M1 and E1 decays of these states produce charmonium states, $^1S_0$ $(0^{-+})$, such as, $\eta_c$, and $^3P_1$  $(1^{++})$ such as, $\chi_{c1}$  respectively. Now, amongst the charmonia below $D\bar{D}$ threshold, the axial, $h_c$  $(^1P_0)$ is the least accessible. We wish to mention that $1^{+-}$ meson state was first detected in $p\bar{p}$ collisions by $R704$ collaboration \cite{r704}. In 1992, E760 reported the observation of the $h_c$ in the $J/\Psi \pi^0$  decay mode, in the reaction, $p\bar{p}$ $\rightarrow$ $ h_c$ $\rightarrow$ $\pi^0 + J/\Psi$ at  $M_{h_c}=3526.2\pm 0.15\pm 0.2$ MeV with $\Gamma_{h_{c}} \leq 1.1$ MeV.\cite{armstrong92}.

In 2005, FNAL E760 \cite{FNAL} analysed two decay modes of $h_c$, the  $\pi^0 J/\Psi$ decay mode, and the $\eta_c \gamma$ decay mode through the reactions, (a) $p\bar{p} \rightarrow  h_c \rightarrow \pi^0 + J/\Psi$; $J/\Psi \rightarrow e^- e^+$; $\pi^0 \rightarrow \gamma \gamma$, and (b) $p\bar{p} \rightarrow  h_c \rightarrow \eta_c \gamma$; $\eta_c \rightarrow \gamma \gamma$, using data for both the runs. They found statistically significant enhancement with mass, $M_{h_c}=3525.8\pm 0.2\pm 0.2 MeV.$, and $\Gamma_{h_c}< 1 MeV.$  The observation of $h_c$ is important, since it provides information on the spin dependence of quark-anti quark interactions. However, the best clue for $h_c(1P)$ came from the CLEO collaboration from isospin violating transition, $e+\bar{e} \rightarrow  \Upsilon (2S) \rightarrow h_c + \pi^0$ \cite{cleo05}. And very  recently, BES III collaboration reported $h_c$ production in the process,  $e^+ e^- \rightarrow \pi^+ \pi^- h_c$ \cite{BESIII}.

The $1^{++}$ mesons are seen in $pp$ collisions. However, not many decays of these mesons are experimentally observed as can be checked from PDG tables \cite{olive14, zyla2020}.

The M1 and E1 transitions of charmonia (that includes axial quarkonia) are quite interesting, and have been recently studied in various models, such as relativistic quark models \cite{eichten08, brambila04}, effective field theory\cite{brambila06,pineda13}, Light-front quark models \cite{vary18,Choi07,shi17}, Lattice QCD \cite{becirevic13,donald12}, Bethe-Salpeter equation \cite{mitra01,karmanov15,he21, wangjhep,bhatnagar20}, and Potential models\cite{deng17}.

 In a recent work\cite{bhatnagar20}, we calculated the radiative M1 decays $1^{--}\rightarrow 0^{-+} \gamma$, and radiative E1 decays involving scalar mesons ($^1 P_1)$ such as, $0^{++}\rightarrow 1^{--} \gamma$, and $1^{--} \rightarrow 0^{++} \gamma$. In the present paper, we focus on the E1 decays involving axial ($1^{+-}$) quarkonia, and the M1 decays of pseudoscalar charmonia, $0^{-+} \rightarrow 1^{--} \gamma$. Thus, in this work, we study E1 radiative transitions involving these axial mesons through processes such as,  $1^{+-} \rightarrow 0^{-+}\gamma$ (such as $h_c\rightarrow \eta_c \gamma$), $0^{-+}-> 1^{+-} \gamma$ (such as $\eta_c(2S) \rightarrow h_c \gamma$), along with M1 transitions, $0^{-+}\rightarrow 1^{--}\gamma$ (such as $\eta_c(2S)\rightarrow J/\Psi \gamma$), which have been studied by some models, for which experimental data\cite{zyla2020,olive14} is available for only some of the transitions. The transitions involving leptonic and radiative decays of axial vector quarkonia would also serve as a test for the  wave functions of these mesons calculated analytically by solving their mass spectral equations\cite{vaishali21} in a recent work.

We wish to mention that decay rates of M1 transitions are much weaker than the rates for E1 transitions. But M1 decay rates are interesting as they allow access to spin-singlet states, that are very difficult to produce. Thus, as regards M1 decays, we study decays, $\eta_c(2S)\rightarrow J/\Psi(1S)\gamma$, and $B_c(2S)\rightarrow B^*_c(1S)\gamma$. $B_c$ meson was discovered in $p\bar{p}$ collisions at $\sqrt{s}$= 1.8 TeV. using CDF detector at Fermilab tevatron. It is the only heavy meson with two heavy quarks with different flavours that forbid their decays into two photons.

We calculate the radiative decay widths of these heavy-light quarkonia for the above mentioned processes in the framework of $4\times 4$ Bethe-Salpeter equation (BSE), which is a fully relativistic approach that incorporates the relativistic effect of quark spins and can also describe internal motion of constituent quarks within the hadron in a relativistically consistent manner, due to its covariant structure\cite{mitra91,hluf16}. Our wave functions satisfy the 3D BSE, which is in turn obtained from 3D reduction of the 4D BSE under Covariant Instantaneous Ansatz (which is a Lorentz-invariant generalization of Instantaneous Approximation), already has relativistic effects. Further, our transition amplitudes also have relativistically covariant form.

The present work, where we make use of two leading Dirac structures in the structure of BS wave functions of $P(0^{-+})$, $V(1^{--})$ and $A^-(1^{+-})$ quarkonia involved in these radiative transitions is more rigorous than our previous work in \cite{bhatnagar20}, where we made use of only the most leading Dirac structure ($\gamma_5$ for P-mesons, $i\gamma.\epsilon$ for V-meson, and $I$ for S-meson) in the BS wave functions of the hadrons involved in the processes.

Our previous studies on mass spectral calculations of heavy-light quarkonia\cite{eshete19, bhatnagar18} were
used to fit the input parameters of our model as $C_0$= 0.69, $\omega_0$= 0.22 GeV, $\Lambda_{QCD}$= 0.250 GeV, and $A_0$= 0.01, with input quark masses
$m_u$= 0.300 GeV, $m_s$= 0.430 GeV,$m_c$= 1.490 GeV, and $m_b$= 4.690 GeV. In the present work on radiative decays, we use these same input parameters to calculate the single photon decay widths for the above processes.

Now, as mentioned in our previous works \cite{bhatnagar18,bhatnagar14,eshete19,hluf16,hluf17}, we are not only interested in studying the mass
spectrum of hadrons, which no doubt is an important element to
study dynamics of hadrons, but also the hadronic wave functions
that play an important role in the calculation of decay constants,
form factors, structure functions etc. for $Q\overline{Q}$, and
$Q\overline{q}$ hadrons. These hadronic Bethe-Salpeter wave
functions were calculated algebraically by us in \cite{hluf16,bhatnagar18, eshete19}. The plots of these wave functions \cite{eshete19} show that they
can provide information not only about the long distance non-perturbative physics, but also act as a bridge between the long distance, and short
distance physics, and are provide us information about the contribution of the short ranged coulomb interactions in the mass spectral calculation of
heavy-light quarkonia. These wave functions can also lead to studies on a number of processes involving
$Q\overline{Q}$, and $Q\overline{q}$ states, and provide a guide for future experiments.

This paper is organized as follows: In section 2, we give the general formulation of the
process, $H\rightarrow H'+\gamma$ in the framework of $4\times 4$ Bethe-Salpeter equation under the covariant instantaneous ansatz. In sections 3, we calculate the single photon decay widths for the processes, $P\rightarrow V\gamma$, In Section 4, we deal with the processs, $A^-\rightarrow P\gamma$. In Section 5, we deal with the process, $P\rightarrow A^-\gamma$, where, $P, A^-$, and $V$ are the pseudoscalar, axial vector, and vector heavy-light quarkonium states. In section 6, we give numerical results and discussions.

\section{Radiative decay process, $H -> H' +\gamma$ in $4\times 4$ BSE under Covariant Instantaneous Ansatz}

The Bethe-Salpeter equation that describes the quark-anti quark bound state of momenta $p_1$ and $p_2$, relative
momentum $q$, and meson momentum $P$ is
\begin{equation}
S_{F}^{-1}(p_{1})\Psi(P,q)S_{F}^{-1}(-p_{2}) =
i\int \frac{d^{4}q''}{(2\pi)^{4}}K(q,q'')\Psi(P,q''),
\end{equation}
where $K(q,q'')$ is the interaction kernel, and $S_{F}^{-1}(\pm p_{1,2})=\pm i{\not}p_{1,2}+ m_{1,2}$ are the quark and antiquark propagators.
We now make use of the Covariant Instantaneous Ansatz (which is a Lorentz-covariant 3D support), where,
$K(q,q'')=K(\widehat{q},\widehat{q}'')$ on the BS kernel, where the BS kernel depends entirely on the variable,
$\widehat{q}_\mu= q_\mu- \frac{q.P}{P^2}P_\mu$ is the component of
internal momentum of the hadron that is orthogonal to the total
hadron momentum, i.e. $\widehat{q}.P=0$, while $\sigma
P_\mu=\frac{q.P}{P^2}P_\mu$ is the component of $q$ longitudinal
to $P$, where the 4-dimensional volume element is,
$d^4q=d^3\widehat{q}Md\sigma$. Now working on the right side of Eq.(1), and making use of the fact that

\begin{equation}
\psi(\hat{q}'')=\frac{i}{2\pi}\int Md\sigma'' \Psi(P,q''),
\end{equation}

and the fact that the longitudinal component of $Md\sigma''$  of $q''$ does not appear in $K(\hat{q}, \hat{q}'')$, carrying out integration over $Md\sigma$ on right side of Eq.(1), we obtain,

\begin{equation}
S_{F}^{-1}(p_{1})\Psi(P,q)S_{F}^{-1}(-p_{2})=\int \frac{d^3\hat{q}''}{(2\pi)^3}K(\hat{q}, \hat{q}'')\psi(\hat{q}'')=\Gamma(\hat{q}),
\end{equation}
where, $\Gamma(\hat{q})$ is the hadron-quark vertex function, and is directly related to the 4D wave function, $\Psi(P,q)$, and one can express the 4D BS wave function $\Psi(P,q)$ in terms of $\Gamma(\hat{q})$ as,

\begin{equation}
 \Psi(P, q)=S_1(p_1)\Gamma(\hat q)S_2(-p_2),
\end{equation}

Further the 4D hadron-quark vertex, that enters into the definition of the 4D BS wave function in the previous equation, can be identified as,
\begin{equation}\label{6a}
 \Gamma(\hat q)=\int\frac{d^3\hat q''}{(2\pi)^3}K(\hat q,\hat q'')\psi(\hat q'').
\end{equation}

Following a sequence of steps outlined in \cite{hluf16}, we get four Salpeter equations which are
effective 3D forms of BSE (Salpeter equations) given below:

\begin{eqnarray}
 &&\nonumber(M-\omega_1-\omega_2)\psi^{++}(\hat{q})=\Lambda_{1}^{+}(\hat{q})\Gamma(\hat{q})\Lambda_{2}^{+}(\hat{q})\\&&
   \nonumber(M+\omega_1+\omega_2)\psi^{--}(\hat{q})=-\Lambda_{1}^{-}(\hat{q})\Gamma(\hat{q})\Lambda_{2}^{-}(\hat{q})\\&&
\nonumber \psi^{+-}(\hat{q})=0.\\&&
 \psi^{-+}(\hat{q})=0\label{fw5}
\end{eqnarray}

Thus, in our framework, a crucial role is played by the component, $\hat{q}_{\mu}$, which is always orthogonal to $P_{\mu}$ and satisfies the unconstrained relation, $\hat{q}.P=0$, regardless of whether $q.P=0$ (i.e. $\sigma=0$), or $q.P\neq 0$ (i.e. $\sigma \neq 0$).

The Lorentz-invariant nature of $\hat{q}^2$ increases the applicability of this framework of Covariant Instantaneous Ansatz all the way from low energy spectra to high energy transition amplitudes. For details, see \cite{bhatnagar20,bhatnagar14}.

The 3D B.S. wave function can be expressed in terms of the projected wave functions as
\begin{equation}\label{bb1}
\psi(\hat q)=\psi^{++}(\hat q)+\psi^{+-}(\hat q)+\psi^{-+}(\hat q)+\psi^{--}(\hat q),
\end{equation}
where
\begin{equation}\label{a7}
 \psi^{\pm\pm}(\hat q)= \Lambda^\pm_{1}(\hat q)\frac{{\not}P}{M}\psi(\hat q)
 \frac{{\not}P}{M}\Lambda^\pm_{2}(\hat q)
\end{equation}
and the projection operators
\begin{equation}\label{a1}
 \Lambda^\pm_{j}(\hat q)=\frac{1}{2\omega_j}\bigg[\frac{{\not}P}{M}\omega_j\pm J(j)(im_j+{\not}\hat q)\bigg],~~~J(j)=(-1)^{j+1},~~j=1,2
\end{equation}
with the relation
\begin{equation}
 \omega^2_j=m_j^2+\hat q^2
\end{equation}

\begin{figure}[ht]
 \centering
\includegraphics[width=20cm,height=8cm]{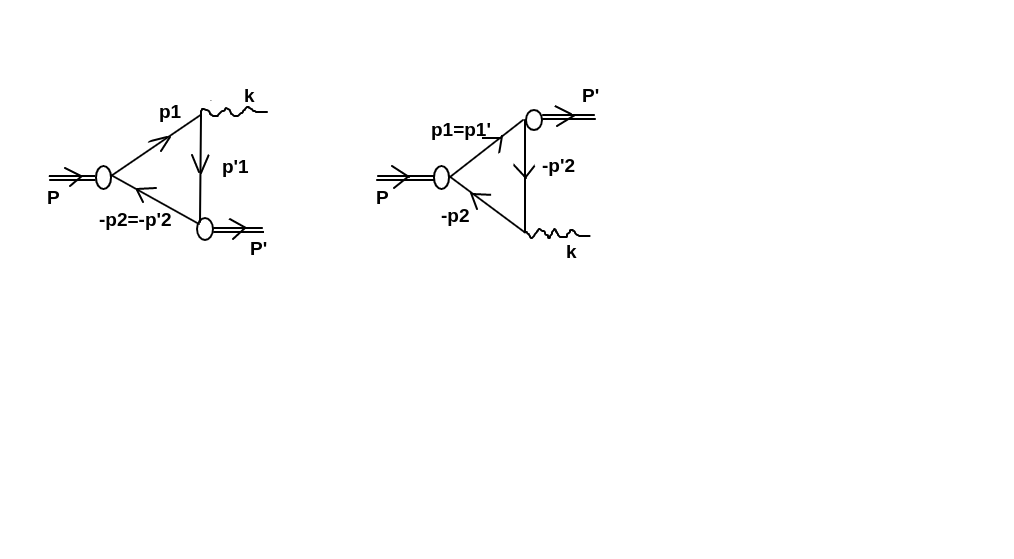}
 \caption{Radiative decays of heavy-light quarkonia}
 \label{fig:1}
\end{figure}

In radiative transitions involving single photon decays, such as $H\rightarrow H'+\gamma$, the
process requires calculation of triangle quark-loop diagram, which involves
two hadron-quark vertices that we attempt in the $4\times 4$ representation of BSE. The single photon decay of $Q\bar{q}$ quarkonia is described by the direct and exchange Feynman diagrams as in Figure \ref{fig:1}.

These processes involve two Lorentz frames, the rest frame of the initial meson, $H$,  and the rest frame of final meson, $H''$. Let $P$, and $q$ be the total momentum and the internal momentum of initial hadron, while $P'$, and $q'$  the corresponding variables of the final hadron. And let $k$, and $\epsilon^{\lambda'}$ be momentum and polarization vectors of emitted photon, while $\epsilon^{\lambda}$ is the polarization vector of final emitted meson. Thus if $p_{1,2}$, and $p'_{1,2}$ are the momenta of the two quarks in initial and final hadron respectively, then, we have, the momentum relations:

\begin{eqnarray}
&&\nonumber P=p_1+p_2;  p_{1,2}=\hat{m}_{1,2}P\pm q\\&&
 P'=p'_1+p'_2;  p'_{1,2}=\hat{m}_{1,2}P'\pm q'
\end{eqnarray}

for initial and final hadrons respectively. From the Feynman diagrams we see that conservation of momentum demands that, $P= P'+k$. Now, for the first diagram, we have the kinematical relations, $p_1=p'_1+k$, and $-p_2=-p'_2$, where $k=P-P'$ is the momentum of the emitted photon. And for the second diagram, we have the corresponding relations, $p_1=p'_1$, and $-p'_2=-p_2 +k$.

Making use of the above equations, the relationship between the internal momenta, $q$, and $q'$ of the initial and final hadrons in terms of the photon momentum, $k$ can be expressed as,

\begin{eqnarray}\label{9}
&&\nonumber q'= q- \hat{m}_2 k,\\&&
q'=q+ \hat{m}_1 k,
\end{eqnarray}

where, the first equation is for Diagram 1, and second equation is for Diagram 2. Here, $\hat{m}_{1,2}=\frac{1}{2}[1\pm\frac{(m^{2}_{1}-m^{2}_{2})}{M^{2}}]$ are the Wightman-Garding definitions\cite{bhatnagar14} of masses of individual quarks, which act like momentum partitioning functions for the two quarks in a hadron.

For initial hadron, its internal momentum has already been decomposed as, $q=(\hat{q},iM\sigma)$, where $\hat{q}$, and $M\sigma$ are defined relative to its external momentum, $P$. Similarly for final meson, we again decompose its internal momentum, $q'$ into two components $q'=(\hat{q}',iM\sigma')$, with $\hat{q}'=q'-\sigma'P$ transverse to initial hadron momentum, $P$, and $\sigma'=\frac{q'.P}{P^2}$, longitudinal to $P$. Thus, $P.\hat{q}'=0$. The relationship between the transverse components of internal momenta of the two hadrons, $\hat{q}$, and $\hat{q}'$ is \cite{bhatnagar20},

\begin{eqnarray}
&&\nonumber \hat{q}'=\hat{q}+\hat{m}_2 \hat{P}'\\&&
\nonumber \hat{q}'=\hat{q}-\hat{m}_1 \hat{P}',\\&&
\hat{P}'=P'-\frac{P'.P}{P^2}P,
\end{eqnarray}

where, the first equation of Eq.(13), holds for Diagram 1, and the second equation holds for Diagram 2, and $\hat{P}'$ is the component of total momentum $P'$ of final hadron transverse to initial hadron momentum, $P$. Here, $\hat{q}'.P=0$ due to $\hat{P'}.P=0$. Now, the kinematics gets simplified in the rest frame of the initial meson, where we have $P=(\overrightarrow{0}, iM)$, while for emitted meson, $P'=(\overrightarrow{P}', iE')$, where $E'=\sqrt{\overrightarrow{P}'^2+M'^2}$, and since the photon momentum can be decomposed as, $k=(\overrightarrow{k},i|\overrightarrow{k}|)$, where $\overrightarrow{k}=-\overrightarrow{P}'$, since final meson and photon would be emitted in opposite directions. Hence we get, $|\overrightarrow{P}'|=|\overrightarrow{k}|=\frac{M^2-M'^2}{2M}$. Thus the energy of the emitted meson can be expressed as, $E'=\frac{M^2+M'^2}{2M}$.

The dot products of momenta of the initial and the emitted meson can be expressed as \cite{bhatnagar20},

\begin{equation}
P'.P=-ME'=-\frac{M^2+M'^2}{2}
\end{equation}

Thus, it can be seen that, $-E'$ acts as the projection of $P'$ along the direction of initial hadron momentum, $P$. Similarly, the relationship between the longitudinal components of internal momenta of the two hadrons in the two diagrams can be worked out as\cite{bhatnagar20},

\begin{eqnarray}\label{13}
&&\nonumber \sigma'= \sigma+\alpha; \\&&
\nonumber \alpha=\hat{m}_{2}\frac{M'^2-M^2}{2M^2} (\Rightarrow  Diagram 1)\\&&
\alpha = -\hat{m}_{1} \frac{M'^2-M^2}{2M^2} (\Rightarrow Diagram 2),
\end{eqnarray}

which is again a consequence of the transversality of $\hat{q}'$ with initial hadron momentum, $P$. Thus, up to Eq.(\ref{13}), the kinematics is the same for all the three processes ($P\rightarrow V\gamma$, $A^-\rightarrow P \gamma$, and $P \rightarrow A^- \gamma$) studied in this work.

\section{M1 Radiative decays, $P\rightarrow V\gamma$}
In this section we study the M1 radiative decay process, $P\rightarrow V\gamma$, having  studied the process, $V\rightarrow P\gamma$ in our previous work\cite{bhatnagar20}. In the present study, we make use of two leading Dirac structures (as in Eq.(35)) in the BS wave functions of the two hadrons involved in the process, in contrast to a single most leading Dirac structure ($\gamma_5$ for P meson, and $i\gamma.\epsilon$ for V meson) used in our previous work \cite{bhatnagar20}. This makes the calculations more involved than with use of the single most leading Dirac stricture in the 4D hadronic BS wave functions. It is to be noted that 4D BS wave function of initial pseudoscalar meson involved in the process is  $\Psi_P(P,q)=S_F(p_1)\Gamma_P(\hat{q})S_F(-p_2)$. Since for transition amplitude calculation, we choose to do calculation in the rest frame of the initial pseudoscalar meson, we write the wave function of the emitted vector meson as,

\begin{equation}
\Psi_{V}(P',q')=S_{F}(p'_{1})\Gamma_V(\hat{q}')S_F(-p'_{2}),
\end{equation}

where as defined earlier, $\hat{q}'=q'-\frac{q'.P}{P^2}P$ is transverse to initial hadron momentum, $P$, where the hadron-quark vertex function $\Gamma_V(\hat{q}')$ for the final meson is,

\begin{equation}\label{6b}
\Gamma_V(\hat q')=\int\frac{d^3\hat q'''}{(2\pi)^3}K(\hat q',\hat q''')\psi(\hat q''').
\end{equation}

Similarly for final meson, the expression for the 3D BS wave function, $\psi(\hat q')$ is expressed in terms of the projection operators as,

\begin{equation}\label{bb2}
\psi(\hat q')=\psi^{++}(\hat q')+\psi^{+-}(\hat q')+\psi^{-+}(\hat q')+\psi^{--}(\hat q')
\end{equation}
where
\begin{eqnarray}\label{a8}
&&\nonumber \psi^{\pm\pm}(\hat q')= \Lambda^\pm_{1}(\hat q')\frac{{\not}P}{M}\psi(\hat q') \frac{{\not}P}{M}\Lambda^\pm_{2}(\hat q')\\&&
\Lambda^\pm_{j}(\hat q')=\frac{1}{2\omega'_j}\bigg[\frac{{\not}P}{M}\omega'_j\pm J(j)(im_j+{\not}\hat q')\bigg],
\end{eqnarray}
with the relation $\omega'^2_{1,2}=m_{1,2}^2+\hat q'^2$.

In Eq.(12), due to relationship between $q'$, and $q$, we can express the electromagnetic transition amplitude of the process as a single integral over $d^4 q = d^3 \hat{q} Md\sigma$ as,

\begin{equation}\label{15}
 M_{fi}=-\int \frac{d^4q}{(2\pi)^4}Tr[e_q\overline{\Psi}_{V}(P', q'){\not}\epsilon^{\lambda'}\Psi_{P}(P,q)
S_F^{-1}(-p_2)+ e_{\overline{Q}}\overline{\Psi}_{V}(P', q')S_F^{-1}(p_1)\Psi_{P}(P, q){\not}\epsilon^{\lambda'}],
\end{equation}

Here, the first term corresponds to the first diagram, where the photon is emitted from the quark, while the second term corresponds to the second diagram where the photon is emitted from the antiquark, and $e_{q}$ , and $e_{Q}$ are the electric charge of quark, and antiquark respectively, while  $\epsilon^{\lambda'}_{\mu}$ is the polarization vector of the emitted photon.

We can express Eq.(20) as,

\begin{eqnarray}\label{17}
&&\nonumber M_{fi}=-e_q\int \frac{d^3 \hat q}{(2\pi)^3}\int\frac{iMd\sigma}{(2\pi )} Tr[\bar{\Gamma}_V(\hat q')S_F(p'_1) {\not}\epsilon' S_F(p_1) \Gamma_P(\hat q) S_F(-p_2)]\\&&
-e_Q\int \frac{d^3 \hat q}{(2\pi)^3}\int\frac{iMd\sigma}{(2\pi )} Tr[\bar{\Gamma}_V(\hat q')S_F(p_1)\Gamma_P(\hat q)S_F(-p_2){\not}\epsilon' S_F(-p'_2)]
\end{eqnarray}

Now, we reduce the above equation to the effective 3D form by integrating over the longitudinal component, $Md\sigma$ over the poles of the propagators, $S_F(p_i)$, that are expressed as\cite{bhatnagar20},

\begin{eqnarray}\label{18}
 &&\nonumber S_F(p_1)=\frac{\Lambda_1^+(\hat q)}{M\sigma+\widehat m_1M-\omega_1+i\epsilon}
 +  \frac{\Lambda_1^-(\hat q)}{M\sigma+\widehat m_1M+\omega_1-i\epsilon},\\&&
\nonumber S_F(-p_2)=\frac{-\Lambda_2^+(\hat q)}{-M\sigma+\widehat m_2M-\omega_2+i\epsilon}
 + \frac{-\Lambda_2^-(\hat q)}{-M\sigma+\widehat m_2M+\omega_2-i\epsilon},\\&&
 S_F(p_1')=\frac{\Lambda_1^+(\hat q')}{M\sigma'+\widehat m_1(-E')-\omega_1'+i\epsilon}
 +  \frac{\Lambda_1^-(\hat q)'}{M\sigma'+\widehat m_1(-E')+\omega_1'-i\epsilon}\\&&
 \nonumber S_F(-p_2')=\frac{-\Lambda_2^+(\hat q')}{-M\sigma'+\widehat m_2(-E')-\omega_2'+i\epsilon}
 +  \frac{-\Lambda_2^-(\hat q)'}{M\sigma'+\widehat m_2(-E')+\omega_2'-i\epsilon}
 \end{eqnarray}

We now put the propagators into Eq.(\ref{17}), and multiplying this equation from the left by the relation, $\frac{\not P}{M}\frac{\not P}{M}=-1= \frac{\not P}{M}(\Lambda^+_2(\hat{q}')+\Lambda^-_2(\hat{q}'))$ \cite{wang06}, the transition amplitude can be expressed as,

\begin{eqnarray}\label{19}
&&\nonumber M^{1}_{fi}=-ie\int \frac{d^3\hat{q}}{(2\pi)^3} [\Omega_1+\Omega_2+\Omega_3+\Omega_4];\\&&
\nonumber \Omega_1=\int\frac{d\sigma}{(2\pi )}\frac{i}{M^3}  Tr\bigg[ \frac{-{\not}P\Lambda_2^+(\hat q')\bar{\Gamma}_P(\hat q')\Lambda_1^+(\hat q'){\not}\epsilon'\Lambda_1^+(\hat q)\Gamma_V(\hat q)\Lambda_2^+(\hat q)}{[\sigma-(-\alpha+\widehat m_1\frac{E'}{M}+\frac{\omega'_1}{M})][\sigma-(-\widehat m_1+\frac{\omega_1}{M})][\sigma-(\widehat m_2-\frac{\omega_2}{M})]} \bigg]\\&&
\nonumber \Omega_2=\int\frac{d\sigma}{(2\pi )}\frac{i}{M^3}  Tr\bigg[ \frac{-{\not}P\Lambda_2^+(\hat q')\bar{\Gamma}_P(\hat q')\Lambda_1^+(\hat q'){\not}\epsilon'\Lambda_1^-(\hat q)\Gamma_V(\hat q)\Lambda_2^-(\hat q)}{[\sigma-(-\alpha+\widehat m_1\frac{E'}{M}+\frac{\omega'_1}{M})][\sigma-(-\widehat m_1-\frac{\omega_1}{M})][\sigma-(\widehat m_2+\frac{\omega_2}{M})]}\bigg];\\&&
\nonumber \Omega_3=\int\frac{d\sigma}{(2\pi )}\frac{i}{M^3}  Tr\bigg[\frac{ -{\not}P\Lambda_2^-(\hat q')\bar{\Gamma}_P(\hat q')\Lambda_1^-(\hat q'){\not}\epsilon'\Lambda_1^+(\hat q)\Gamma_V(\hat q)\Lambda_2^+(\hat q)}{[\sigma-(-\alpha+\widehat m_1\frac{E'}{M}-\frac{\omega'_1}{M})][\sigma-(-\widehat m_1+\frac{\omega_1}{M})][\sigma-(\widehat m_2-\frac{\omega_2}{M})]}\bigg]\\&&
\Omega_4=\int\frac{d\sigma}{(2\pi )}\frac{i}{M^3}  Tr\bigg[\frac{-{\not}P\Lambda_2^+(\hat q')\bar{\Gamma}_P(\hat q')\Lambda_1^-(\hat q'){\not}\epsilon'\Lambda_1^-(\hat q)\Gamma_V(\hat q)\Lambda_2^-(\hat q)}{[\sigma-(-\alpha+\widehat m_1\frac{E'}{M}-\frac{\omega'_1}{M})][\sigma-(-\widehat m_1-\frac{\omega_1}{M})][\sigma-(\widehat m_2+\frac{\omega_2}{M})]}\bigg],
\end{eqnarray}

The contour integrations over $Md\sigma$ are performed over each of the four terms taking into account the pole positions in the complex $\sigma$ plane:

\begin{eqnarray}
&&\nonumber \sigma^{\pm}_{3} =-\alpha+\hat{m}_{1}\frac{E'}{M}\mp \frac{\omega'_{1}}{M}\pm i\epsilon\\&&
\nonumber \sigma^{\pm}_{1} =-\hat{m}_{1}\mp \frac{\omega_{1}}{M}\pm i\epsilon\\&&
\sigma^{\pm}_{2} = \hat{m}_{2} \mp \frac{\omega_2}{M}\pm i\epsilon.
\end{eqnarray}

For the second diagram, we can write the amplitude, $M^{2}_{fi}$ as,

\begin{eqnarray}\label{20}
&&\nonumber M^{2}_{fi}=-e_Q\int \frac{d^3\hat{q}}{(2\pi)^3} [\Omega'_1+\Omega'_2+\Omega'_3+\Omega'_4];\\&&
\nonumber \Omega'_1=\int\frac{d\sigma}{(2\pi )}\frac{i}{M^3}  Tr\bigg[ \frac{-\Lambda_2^+(\hat q')\bar{\Gamma}_V(\hat q')\Lambda_1^+(\hat q'){\not}P\Lambda_1^+(\hat q)\Gamma_P(\hat q)\Lambda_2^+(\hat q){\not}\epsilon'}{[\sigma-(-\alpha-\widehat m_2\frac{E'}{M}-\frac{\omega'_2}{M})][\sigma-(-\widehat m_1+\frac{\omega_1}{M})][\sigma-(\widehat m_2-\frac{\omega_2}{M})]} \bigg]\\&&
\nonumber \Omega'_2=\int\frac{d\sigma}{(2\pi )}\frac{i}{M^3}  Tr\bigg[ \frac{-\Lambda_2^-(\hat q')\bar{\Gamma}_V(\hat q')\Lambda_1^-(\hat q'){\not}P\Lambda_1^+(\hat q)\Gamma_P(\hat q)\Lambda_2^+(\hat q){\not}\epsilon'}{[\sigma-(-\alpha-\widehat m_2\frac{E'}{M}+\frac{\omega'_2}{M})][\sigma-(-\widehat m_1+\frac{\omega_1}{M})][\sigma-(\widehat m_2-\frac{\omega_2}{M})]}\bigg];\\&&
\nonumber \Omega'_3=\int\frac{d\sigma}{(2\pi )}\frac{i}{M^3}  Tr\bigg[\frac{ -\Lambda_2^+(\hat q')\bar{\Gamma}_V(\hat q')\Lambda_1^+(\hat q'){\not}P \Lambda_1^+(\hat q)\Gamma_P(\hat q)\Lambda_2^-(\hat q){\not}\epsilon'}{[\sigma-(-\alpha-\widehat m_2\frac{E'}{M}-\frac{\omega'_2}{M})][\sigma-(-\widehat m_1-\frac{\omega_1}{M})][\sigma-(\widehat m_2+\frac{\omega_2}{M})]}\bigg]\\&&
\Omega'_4=\int\frac{d\sigma}{(2\pi )}\frac{i}{M^3}  Tr\bigg[\frac{-\Lambda_2^-(\hat q')\bar{\Gamma}_V(\hat q')\Lambda_1^-(\hat q'){\not}P\Lambda_1^-(\hat q')\Gamma_P(\hat q)\Lambda_2^-(\hat q){\not}\epsilon'}{[\sigma-(-\alpha+\widehat m_2\frac{E'}{M}+\frac{\omega'_2}{M})][\sigma-(-\widehat m_1-\frac{\omega_1}{M})][\sigma-(\widehat m_2+\frac{\omega_2}{M})]}\bigg],
\end{eqnarray}

where the rest of the terms are anticipated to be zero on account of 3D Salpeter equations. The contour integrations over $Md\sigma$ are performed over each of the four terms taking into account the pole positions in the complex $\sigma$ plane:

\begin{eqnarray}
&&\nonumber \sigma^{\pm}_{3} =-\alpha-\hat{m}_{2}\frac{E'}{M}\mp \frac{\omega'_{2}}{M}\pm i\epsilon\\&&
\nonumber \sigma^{\pm}_{1} =-\hat{m}_{1}\pm \frac{\omega_{1}}{M}\pm i\epsilon\\&&
\sigma^{\pm}_{2} = \hat{m}_{2} \mp \frac{\omega_2}{M}\pm i\epsilon.
\end{eqnarray}

The contour integral over each of the four terms can be performed by closing the contour either above or below the real axis in the complex $\sigma$- plane.

We now make use of the Salpeter equations in variable, $\hat{q}$ in Eq.(6), and the Salpeter equations in variable $\hat{q}'$ given below. It is to be noted that the Salpeter equations in $\hat{q}'$ involve $-E'=\frac{P.P'}{M}$, which is the projection of $P'$ along the direction of initial momentum, $P$, and are given as \cite{wang06},

\begin{eqnarray}
 &&\nonumber(-E'-\omega'_1-\omega'_2)\psi^{++}(\hat{q'})=\Lambda_{1}^{+}(\hat{q}')\Gamma(\hat{q}')\Lambda_{2}^{+}(\hat{q}')\\&&
   \nonumber(-E'+\omega'_1+\omega'_2)\psi^{--}(\hat{q}')=-\Lambda_{1}^{-}(\hat{q}')\Gamma(\hat{q}')\Lambda_{2}^{-}(\hat{q}')\\&&
\nonumber \psi^{+-}(\hat{q}')=0,\\&&
 \psi^{-+}(\hat{q}')=0\label{fw9}
\end{eqnarray}

It can be verified that the results of each of these four integrals, $\Omega_1,...,\Omega_4$ whether we close the contour above or below the real $\sigma$-axis comes out to be the same, thereby validating the correctness of the formalism employed. These results of integrals over $d\sigma$ in $\Omega_1,...,\Omega_4$, are given as, $\alpha_1,..., \alpha_4$ in Eqs.(32).

This leads to the expression for effective 3D form of transition amplitude, $M^{1}_{fi}$ under Covariant Instantaneous Ansatz for Diagram 1, as,

\begin{multline}\label{7f}
 M^{1}_{fi}=-ie\int \frac{d^3 \hat q}{(2\pi)^3} \frac{1}{M^2}Tr\bigg[ \alpha_1 {\not}P\overline{\psi}_P^{++}(\hat q'){\not}\epsilon'\psi_V^{++}(\hat q)
 + \alpha_2 {\not}P\overline{\psi}_P^{++}(\hat q'){\not}\epsilon'\psi_V^{--}(\hat q)\\
 +\alpha_3 {\not}P\overline{\psi}_P^{--}(\hat q'){\not}\epsilon'\psi_V^{++}(\hat q)
 + \alpha_4 {\not}P\overline{\psi}_P^{--}(\hat q'){\not}\epsilon'\psi_V^{--}(\hat q)\bigg]
 \end{multline}

Here, the results of contour integrals over $d\sigma$ are given as $\alpha_1,...,\alpha_4$:

\begin{align}
 \nonumber \alpha_1&=\frac{[-E'-\omega'_1-\omega'_2]}{[\alpha-\widehat m_1\frac{E'}{M}+\widehat m_2-\frac{1}{M}(\omega'_1+\omega_2)]}\\
 \nonumber \alpha_2&=\frac{-[-E'-\omega'_1-\omega'_2]}{[\alpha-\widehat m_1(\frac{E'}{M}-1)-\frac{1}{M}(\omega_1+\omega'_1)]}\\
 \nonumber \alpha_3&=\frac{[-E'+\omega'_1+\omega'_2]}{[\alpha-\widehat m_1(\frac{E'}{M}-1)+\frac{1}{M}(\omega_1+\omega'_1)]}\\
  \alpha_4&=\frac{-[-E'+\omega'_1+\omega'_2]}{[\alpha-\widehat m_1\frac{E'}{M}+\widehat m_2+\frac{1}{M}(\omega'_1+\omega_2)]},
\end{align}

and the projected wave functions, $\psi^{\pm \pm}$ being taken from the 3D Salpeter equations \cite{eshete19} derived earlier, which for initial meson in internal variable $\hat{q}$ are given in Eq.(6).

[It is to be noted that the factors $(M\pm \omega_1 \pm \omega_2)$ that were also present in the numerators of $\alpha$'s in Eq.(29) as a result of the first two Salpeter equations in variable, $\hat{q}$ in Eqs. (6), get cancelled from the corresponding factors (in denominator) resulting from contour integrals over $d\sigma$, while the numerators of $\alpha_1,..., \alpha_4$ come from the Salpeter equations in variable, $\hat{q}'$ in Eq.(27).]

Similarly, the expression for effective 3D form of transition amplitude, $M^{2}_{fi}$ under Covariant Instantaneous Ansatz can be expressed as,

\begin{multline}\label{10f}
 M^{2}_{fi}=-e_Q\int \frac{d^3 \hat q}{(2\pi)^3} \frac{1}{M^2}Tr\bigg[ \alpha'_1 \overline{\psi}_P^{++}(\hat q'){\not}P\psi_V^{++}(\hat q){\not}\epsilon'
 + \alpha'_2 \overline{\psi}_P^{++}(\hat q'){\not}P\psi_V^{--}(\hat q){\not}\epsilon'\\
 +\alpha'_3 \overline{\psi}_P^{--}(\hat q'){\not}P\psi_V^{++}(\hat q){\not}\epsilon'
 + \alpha'_4 \overline{\psi}_P^{--}(\hat q'){\not}P\psi_V^{--}(\hat q){\not}\epsilon'\bigg]
 \end{multline}

Here, the results of contour integrals over $d\sigma$ are given as $\alpha'_1,...,\alpha'_4$:

\begin{align}
  \nonumber \alpha'_1=\frac{[-E'-\omega'_1-\omega'_2]}{[\alpha-\hat m_1+\hat m_2 \frac{E'}{M}+\frac{1}{M}(\omega_1+\omega'_2)]}\\
 \nonumber\alpha'_2=\frac{-[-E'-\omega'_1-\omega'_2]}{[\alpha+\hat m_2(1+ \frac{E'}{M})-\frac{1}{M}(\omega_2+\omega'_2)]}\\
 \alpha'_3=\frac{[-E'+\omega'_1+\omega'_2]}{[\alpha+\hat m_2(1+ \frac{E'}{M})+\frac{1}{M}(\omega_2+\omega'_2)]}\\
 \nonumber \alpha'_4=\frac{-[-E'+\omega'_1+\omega'_2]}{[\alpha-\hat m_1+\hat m_2 \frac{E'}{M}-\frac{1}{M}(\omega_1+\omega'_2)]},
\end{align}

Thus, we make use of the generalized method for handling quark-triangle diagrams with two hadron -quark vertices in the framework of $4\times 4$ BSE under Covariant Instantaneous Ansatz described in \cite{bhatnagar20}, by expressing the transition amplitude, $M_{fi}$ as a linear superposition of terms involving all possible combinations of $++$, and $--$ components of Salpeter wave functions of final and initial hadrons through  $++++$, $----$ , $++--$, and $--++$, with each of the four terms being associated with a coefficient, $\alpha_i (i=1,...,4)$, which is the result of pole integration in the complex $\sigma$-plane, which should be a feature of relativistic frameworks.

Now, to calculate the amplitude, $M_{fi}$ for the process, we need the $++$, and $--$ components, $\psi_V^{\pm \pm}(\hat{q'})$ for vector and $\psi_P^{\pm \pm}(\hat{q})$ for pseudoscalar mesons, given as,

\begin{eqnarray}
&&\nonumber  \psi_P^{\pm\pm}(\hat q)= \Lambda^{\pm}_{1}(\hat q)\frac{{\not}P}{M}\psi_P(\hat q)
 \frac{{\not}P}{M}\Lambda^{\pm}_{2}(\hat q),\\&&
\psi_V^{\pm\pm}(\hat q')= \Lambda^\pm_{1}(\hat q')\frac{{\not}P}{M}\psi_V(\hat q')
 \frac{{\not}P}{M}\Lambda^\pm_{2}(\hat q'),
\end{eqnarray}

to calculate which we need the 3D wave functions, $\psi_P(\hat q)$, and $\psi_V(\hat q')$. To derive these, we start with the general 4D decomposition of BS
wave functions \cite{smith69,alkofer}. Using 3D decomposition under Covariant Instantaneous Ansatz, the wave function of vector mesons of dimensionality, $M$ can be written as \cite{bhatnagar18,hluf16,bhatnagar20},

\begin{equation}\label{wf5}
 \psi^V(\hat q)=iM{\not}\epsilon\chi_1(\hat q)+{\not}\epsilon{\not}P\chi_2(\hat q)+[{\not}\epsilon{\not}\hat q-\hat q.\epsilon]\chi_3(\hat q)
 -i[{\not}P{\not}\epsilon{\not}\hat q+\hat q.\epsilon{\not}P]\frac{1}{M}\chi_4(\hat q)+(\hat q.\epsilon)\chi_5(\hat q)-i
 \hat q.\epsilon\frac{{\not}P}{M}\chi_6(\hat q),
\end{equation}

where $\epsilon^{\lambda}$ is the vector meson polarization vector, while for a pseudoscalar meson, the 3D wave function with dimensionality, $M$ can
be written as \cite{bhatnagar18,hluf16,bhatnagar20},

\begin{equation}\label{wf2}
 \psi^P(\hat q)=N_{P}[M\phi_1(\hat q)-i{\not}P\phi_2(\hat q)+i{\not}\hat q\phi_3(\hat q)+\frac{{\not}P{\not}\hat q}{M}\phi_4(\hat q)]\gamma_5.
\end{equation}

\bigskip

However, in calculation of hadronic observables, it was noticed that some Dirac covariants \cite{alkofer} in structure of hadronic BS wave function covariants contribute much more than others. And in accordance with a naive power counting rule in \cite{bhatnagar06,bhatnagar09, bhatnagar14}, for pseudoscalar mesons, one could classify Dirac structures, $M\gamma_5$, and ${\not}P \gamma_5$ associated with amplitudes $\phi_1$, and $\phi_2$ respectively as leading, while those with $\phi_3$, and $\phi_4$ as sub-leading. A similar behaviour was observed in case of vector mesons\cite{bhatnagar06,bhatnagar14}, where Dirac structures, $M{\not}\epsilon$ and ${\not}\epsilon {\not}P$ associated with $\chi_1$ and $\chi_2$ respectively are leading\cite{bhatnagar06,bhatnagar14}, while those associated with $\chi_3,..., \chi_6$ are sub-leading. A similar observation about the most leading Dirac structures from all the Dirac structures was made by \cite{munczek92,sauli13}. Thus to simplify the algebra, we take 3D wave functions with these two leading Dirac structures (the present work with two leading Dirac structures is more involved than our previous work \cite{bhatnagar20}, where we took only the most leading Dirac structure in hadronic BS wave functions of P, V, and S mesons\cite{bhatnagar20}),

\begin{equation}\label{wf44}
\begin{split}
  \psi_P(\hat q)=N_P\bigg[M\gamma_5-i{\not}P\gamma_5\bigg]\phi_P(\hat{q})\\
 \psi_V(\hat q)=N_V\bigg[iM{\not}\epsilon+{\not}\epsilon {\not}P\bigg]\phi_V(\hat{q}),
 \end{split}
\end{equation}

with the 3D radial wave functions, $\phi_P(\hat{q})$, and $\phi_V(\hat{q})$, obtained as solutions of mass spectral equations\cite{hluf16,bhatnagar18,eshete19} of pseudoscalar, and vector quarkonia respectively, that were obtained from 3D Salpeter equations being,

\begin{equation}\label{25}
\begin{split}
   \phi_{P,V}(1S,\hat q)&=\frac{1}{\pi^{3/4}}\frac{1}{\beta_{P,V}^{3/2}}e^{-\frac{\hat q^2}{2\beta_{P,V}^2}}\\
 \phi_{P,V}(2S,\hat q)&=\sqrt{\frac{3}{2}}\frac{1}{\pi^{3/4}}\frac{1}{\beta_{P,V}^{3/2}}
  \bigg(1-\frac{2\hat q^2}{3\beta_{P,V}^2}\bigg)e^{-\frac{\hat q^2}{2\beta_{P,V}^2}}\\
  \phi_V(1D,\hat q)&=\sqrt{\frac{4}{15}}\frac{1}{\pi^{3/4}}\frac{1}{\beta_V^{7/2}}\hat q^2e^{-\frac{\hat q^2}{2\beta_V^2}}\\
\phi_{P,V}(3S,\hat q)&=\sqrt{\frac{15}{8}}\frac{1}{\pi^{3/4}}\frac{1}{\beta_{P,V}^{3/2}}
     \bigg(1-\frac{4\hat q^2}{3\beta_{P,V}^2}+\frac{4\hat q^4}{15\beta_{P,V}^4}\bigg)e^{-\frac{\hat q^2}{2\beta_{P,V}^2}},
 \end{split}
\end{equation}
where $\beta_{P,V}$ \cite{eshete19} are the inverse range parameters.
\bigskip

These radial wave functions were recently used for calculations of some M1 and E1 transitions in \cite{bhatnagar20}. We have made use of the 3D Salpeter equations in Eq.(6), that depend on the variable $\hat{q}^2$, that is Lorentz-invariant, and is a four-scalar, whose validity extends over the entire 4D space, while also keeping contact with the surface, $P.q=0$ (hadron rest frame). Now, our  mass spectrum, and the 3D wave functions $\phi(\hat{q})$ in Eq.(36) and (49) (please see \cite{hluf16}) were calculated from Salpeter equations in Eq.(6) in the rest frame of the hadron.

Here, $\hat{q}'^2$ is expressed as \cite{bhatnagar20},

\begin{equation}
\hat{q}'^2=\hat{q}^2+2\hat{m}_2 \frac{(M^2-M'^2)}{2M}|\hat{q}|+\hat{m}_2^2\frac{(M^2-M'^2)^2}{4M^2}
\end{equation}
where,$|\hat{q}|$ is the length of the 3-D vector, $\hat{q}$, defined as $|\hat{q}| =\sqrt{\hat{q}^2}=\sqrt{q^2-(q.P)^2/P^2}$, and is a Lorentz-invariant variable Ref.\cite{bhatnagar20}.

The Bethe-Salpeter normalizers, $N_P$, and $N_V$ are obtained through current conservation condition,
\begin{equation}
2iP_{\mu}=\int \frac{d^4 q}{(2\pi)^4} Tr\bigg[\overline{\psi}(P,q)[\frac{\partial}{\partial P_{\mu}}S_F^{-1}(p_1)]\psi(P,q)S_F^{-1}(-p_2)\bigg] +(1 \leftrightarrow 2),
\end{equation}

Substituting the 3D BS wave function of pseudoscalar meson, the $++$ and $--$ components of the 3D BS wave function of pseudoscalar meson are given in Eq.(\ref{d5}) of Appendix. Similarly, the corresponding adjoint wave functions are given in Eq.(\ref{d6}) of Appendix. Similarly, the corresponding $++$, and $--$ components of vector mesons are again given as Eq.(\ref{d71}), and Eq.(\ref{d72}) of Appendix. We can similarly construct the $++$, and $- -$ components of axial meson wave functions.

The transition amplitude, $M_{fi}$ is expressed as,

\begin{equation}
M_{fi}=F_{PV}~\epsilon_{\mu\nu\alpha\beta}~P_\mu \epsilon^{\lambda'}_\nu \epsilon^{\lambda}_\alpha P'_\beta ,
\end{equation}
where the antisymmetric tensor, $\epsilon_{\mu\nu\alpha\beta}$ ensures its gauge invariance. Here, $F_{PV}$ is the transition form factor for $P\rightarrow V\gamma$, with expression,

\begin{equation}\label{11f}
F_{PV}=F^1_{PV}+F^2_{PV}=N_PN_V\frac{1}{M^2}\int \frac{d^3\hat q}{(2\pi)^3} \frac{\phi_P(\hat q)\phi_V(\hat q)}{16\omega_1\omega_2\omega'_1\omega'_2}\bigg[e_q [TR_1] + e_Q [TR_2]\bigg] ,
\end{equation}

where the expressions for $F_{PV}^1$ and $F_{PV}^1$ are given in Eqs.(80)-(81) in the Appendix, and correspond to contributions from the two diagrams, which is a more accurate treatment. Here $[TR_1]$, and $[TR_2]$ involve trace over gamma matrices. The above expression corresponds to $F_{PV}(k^2=0)$, that corresponds to emission of a real photon.

The kinematical relation connecting $\hat{q}'^2$, with $\hat{q}^2$, is given in Eq.(37). To calculate the decay widths, we need to calculate the spin averaged amplitude square, $|\overline{M}_{fi}|^2$, where
$|\overline{M}_{fi}|^2=\sum_{\lambda,\lambda'}|{M}_{fi}|^2$, where we sum over the final polarization states $\lambda$ of V-meson, and $\lambda'$ of photon. We make use of the normalizations,
$\Sigma_{\lambda} \epsilon_{\mu}^{\lambda}\epsilon_{\nu}^{\lambda}=\frac{1}{3}(\delta_{\mu\nu}+\frac{P_{\mu}P_{\nu}}{M^2})$ for vector meson, and
$\Sigma_{\lambda'} \epsilon_{\mu}^{\lambda'}\epsilon_{\nu}^{\lambda'}=\delta_{\mu\nu}$, for the emitted photon, with $M_{fi}$ taken from Eq.(39).
Thus, we write,

\begin{equation}
 |\overline{M_{fi}}|^2=-2e^2[M^2M'^2-(P.P')^2]~|F_{PV}|^2
\end{equation}

In the above equation, we evaluate $P.P'=-ME'$ in the rest frame of initial vector meson, where $E'=\sqrt{\overrightarrow{P}'^2+M'^2}$ is the energy of the final pseudoscalar meson, giving,$P.P'=-(\frac{M^2+M'^2}{2})$. Thus, $|\overline{M}_{fi}|^2$ can be expressed as,

\begin{equation}
|\overline{M}_{fi}|^2=e^2\frac{(M^2-M'^2)^2}{2}|F_{PV}|^2.
\end{equation}

The decay width of the process ($P\rightarrow V\gamma$) in the rest frame of the initial meson is expressed as
\begin{equation}
 \Gamma_{P\rightarrow V\gamma}=\frac{|\overline{M}_{fi}|^2}{8\pi M^2}|\overrightarrow{P'}|,
\end{equation}

where we make use of the fact that modulus of the momentum of the emitted vector meson can be expressed in terms of masses of particles as,
$|\overrightarrow{P'}|=|\overrightarrow{k}|=\omega_k=\frac{1}{2M} (M^2-M'^2)$, where, $\omega_k$ is the kinematically allowed energy of the emitted photon. Thus, $\Gamma$ in turn can be expressed as:

\begin{equation}
\Gamma=\alpha_{e.m.}|F_{PV}|^2\omega_k^3.
\end{equation}

The M1 decay widths for the processes studied are given in Table 2, for the meson masses given in Table 1, that were evaluated in our previous works\cite{eshete19,vaishali21}

\begin{table}[ht]
\begin{center}
\begin{tabular}{p{1.2cm}  p{2.7cm}  p{2.6cm} p{2.6cm} p{2.8cm} p{3cm}  }
\hline\hline
 &\footnotesize{BSE-CIA}&\small Expt.\cite{zyla2020}& BSE &\small Ref.\cite{vinodkumar14,burch10,kawanai15,davies96} &\small RQM  \\
\hline
\small $M_{h_c(1P_1)}$ &3.535 & 3.525$\pm$0.00001 &3.5244\cite{wang06} &3.5059\cite{kawanai15} &3.525\cite{souza17}\\
\small $M_{h_c(2P_1)}$&3.743 & 3.888$\pm$0.0025&3.9358\cite{wang06}&&3.927\cite{souza17}\\

\small$M_{\eta_c(1S_0)}$&3.0004 & 2.9839$\pm$0.0005& & 3.292\cite{burch10} & 2.981\cite{ebert13} \\

\small$M_{\eta_c(2S_0)}$&3.5934 &3.6376$\pm$0.0012 & &4.240\cite{burch10} &3.635\cite{ebert13}\\
\small$M_{D(2S_0)}$&2.5288 &     & &2.5235\cite{vinodkumar14} &2.581\cite{ebert13}\\
\small$M_{D^*(1S_0)}$&2.0221 &2.010$\pm$0.00005 & &2.0104\cite{vinodkumar14} &2.010\cite{ebert13}\\
\small$M_{D_s(2S_0)}$&2.6358 &           & &2.6333\cite{vinodkumar14} &2.6888\cite{ebert13}\\
\small$M_{D_s^*(1S_0)}$&2.0221 &2.010$\pm$0.0004 & &2.1123\cite{vinodkumar14} &2.111\cite{ebert13}\\
\small$M_{J/\Psi(1S)}$& 3.0970& 3.0969$\pm$0.0000025&  &3.099\cite{kawanai15}&3.096\cite{ebert03}\\
\small$M_{B^*_c(1S_1)}$&6.3514 & & &6.321$\pm$0.020\cite{davies96}& 6.332\cite{ebert03}\\
\small$M_{B_c(2S_0)}$& 6.7241&    & & 6.960$\pm$0.080\cite{davies96}&6.835\cite{ebert03}\\
\hline \hline
\end{tabular}
\end{center}
\caption{Masss spectra of ground and excited states of axial
($1^{+-}$), pseudoscalar ($0^{-+}$), and vector ($1^{--}$) quarkonia (in GeV) in BSE-CIA used in the transitions studied in this work, along with data and results of other models}
\end{table}

\begin{table}
\begin{center}
\begin{tabular}{p{3cm} p{1.8cm} p{2.8cm} p{3.1cm} p{2.2cm} p{1.8cm} p{2.6cm} }
  \hline\hline
 &$\beta$ &BSE-CIA &  Expt.   &  RQM    & PM  & RQM   \\
   \hline
 $\Gamma_{\eta_c(2S_0)\rightarrow J/\Psi(1S_1)\gamma}$ &0.200 & 0.516 & $<$ 158 $^{+44.8}_{-40.6}$\cite{zyla2020}& 0.700\cite{ebert03}& 2.29\cite{deng12}& 5.6\cite{godfrey05}\\
 $\Gamma_{D(2S_0)\rightarrow D^*(1S_1)\gamma}$&0.180 & 13.953 & && 8.594\cite{rai17}& \\
 $\Gamma_{D_s(2S_0)\rightarrow D_s^*(1S_1)\gamma}$&0.185 & 6.839 & && & \\
 $\Gamma_{B_c(2S_0)\rightarrow B_c^*(1S_1)\gamma}$&0.460 &  0.1016 &  &0.096\cite{kiselev95} &0.093\cite{eichten94}&0.488\cite{ebert03} \\
\hline\hline
\end{tabular}
\end{center}
\caption{Radiative decay widths of heavy-light mesons (in Kev) for M1 transitions in BSE, along with experimental data  and results of other models.}
\end{table}

We now calculate the radiative decay widths for the process, $A^-\rightarrow P+\gamma$ in the next section.

\section{Radiative decays of heavy-light quarkonia through $A^-\rightarrow P\gamma$}
The E1 transitions are characterized by $|\Delta L|=1$. Thus in these transitions, there is change in parity between the initial and final hadronic states. The scattering amplitude of the decay process $A^-\rightarrow P\gamma$ corresponding to Diagram 1, can be written as,

\begin{multline}\label{7f1}
 M^{1}_{fi}=-ie\int \frac{d^3 \hat q}{(2\pi)^3} \frac{1}{M^2}Tr\bigg[ \alpha_1 {\not}P\overline{\psi}_P^{++}(\hat q'){\not}\epsilon'\psi_A^{++}(\hat q)
 + \alpha_2 {\not}P\overline{\psi}_P^{++}(\hat q'){\not}\epsilon'\psi_A^{--}(\hat q)\\
 +\alpha_3 {\not}P\overline{\psi}_P^{--}(\hat q'){\not}\epsilon'\psi_A^{++}(\hat q)
 + \alpha_4 {\not}P\overline{\psi}_P^{--}(\hat q'){\not}\epsilon'\psi_A^{--}(\hat q)\bigg]
 \end{multline}

Here, the results of contour integrals over $d\sigma$ are given as $\alpha_1,...,\alpha_4$ in Eq.(29). Similarly the amplitude for Diagram 2 can be written as,

\begin{multline}\label{8f}
 M^{2}_{fi}=-e_Q\int \frac{d^3 \hat q}{(2\pi)^3} \frac{1}{M^2}Tr\bigg[ \alpha'_1 \overline{\psi}_P^{++}(\hat q'){\not}P\psi_A^{++}(\hat q){\not}\epsilon'
 + \alpha'_2 \overline{\psi}_P^{++}(\hat q'){\not}P\psi_A^{--}(\hat q){\not}\epsilon'\\
 +\alpha'_3 \overline{\psi}_P^{--}(\hat q'){\not}P\psi_A^{++}(\hat q){\not}\epsilon'
 + \alpha'_4 \overline{\psi}_P^{--}(\hat q'){\not}P\psi_A^{--}(\hat q){\not}\epsilon'\bigg]
 \end{multline}

And, the results of contour integrals over $d\sigma$ are given as $\alpha'_1,...,\alpha'_4$ in Eq.(31). After the 3D reduction of the 4D BS wave function of scalar meson under CIA, we express the 3D BS wave function with dimensionality $M$ as

\begin{equation}\label{uw2}
 \psi_{A^-}(\hat q)=\gamma_5 (\epsilon.\hat{q})\bigg[g_1(\hat{q})+i\frac{{\not}P}{M}g_2(\hat{q})-i\frac{{\not} \hat{q}}{M}g_3(\hat{q})+2\frac{{\not} P{\not} \hat{q}}{M^2}g_4(\hat{q})\bigg].
 \end{equation}

Making use of the fact that the leading order Dirac structures in axial meson BS wave function in accordance with the power counting rule proposed in \cite{bhatnagar06, bhatnagar09, bhatnagar14} are $\gamma_5 \epsilon.\hat{q}$, and $i\gamma_5 \epsilon.\hat{q}\frac{{\not}P}{M}$, and making
use of \cite{bhatnagar18}, we express the 3D axial meson BS wave function of dimensionality, $M$ as,

\begin{equation}\label{uw1}
 \psi_{A^-}(\hat q)=\gamma_5 (\epsilon.\hat{q})\bigg[1+i\frac{{\not}P}{M}\bigg]\phi_{A^-}(\hat{q})
 \end{equation}

where $\phi_{A^-}(\hat{q})$ is the spatial part of this wave function, whose analytic forms obtained by power series solutions of 3D mass spectral equations (derived from 3D Salpeter equations in Eq.(6)), in the variable $\hat{q}$ (which is in fact $|\hat{q}|$) for P-wave meson in its own rest frame, calculated in \cite{bhatnagar18} are

\begin{equation}\label{wv1}
\begin{split}
 \phi_A(1P,\hat q)&=\sqrt{\frac{2}{3}}\frac{1}{\pi^{3/4}}\frac{1}{\beta_A^{5/2}} \hat q e^{-\frac{\hat q^2}{2\beta_A^2}}\\
 \phi_A(2P,\hat q)&=\sqrt{\frac{5}{3}}\frac{1}{\pi^{3/4}}\frac{1}{\beta_A^{5/2}}
  \hat q\bigg(1-\frac{2\hat q^2}{5\beta_A^2}\bigg)e^{-\frac{\hat q^2}{2\beta_A^2}}\\
    \phi_A(3P,\hat q)&=\sqrt{\frac{35}{12}}\frac{1}{\pi^{3/4}}\frac{1}{\beta_A^{5/2}}
 \hat q\bigg(1-\frac{4\hat q^2}{5\beta_A^2}+\frac{4\hat q^4}{35\beta_A^4}\bigg)e^{-\frac{\hat q^2}{2\beta_A^2}}\\
   \phi_A(4P,\hat q)&=\sqrt{\frac{35}{8}}\frac{1}{\pi^{3/4}}\frac{1}{\beta_A^{5/2}}
 \hat q\bigg(1-\frac{6\hat q^2}{5\beta_A^2}+\frac{12\hat q^4}{35\beta_A^4}-\frac{8\hat q^6}{315\beta_A^6}\bigg)e^{-\frac{\hat q^2}{2\beta_A^2}},
\end{split}
\end{equation}

These wave functions in Eq.(49) involve even powers of $\hat{q}$, along with odd power, $\hat{q}$.  Here $\hat{q}=|\hat{q}|$ as explained above is the length of the 3D vector $\hat{q}$, and is expressed as $|\hat{q}| = \sqrt{q^2-(q.P)^2/P^2}$, and is a Lorentz-invariant quantity\cite{bhatnagar20}, along with even powers of $\hat{q}$, such as $\hat{q}^2, \hat{q}^4$,... etc. which are again Lorentz-invariant. While for S-wave mesons, the wave functions are only functions of even powers of $\hat{q}$. Thus when P-wave or S-wave mesons are in the final state, their wave functions after Lorentz transformation, would  involve the variables, $|\hat{q}’|$, and/or even powers of $\hat{q}'$.  We express $\hat{q}'^2$ in terms of $\hat{q}^2$ directly through Eq.(37), that connects  $\hat{q}'^2$ with $\hat{q}^2$, while odd power, $|\hat{q'}|$ is expressed as, $|\hat{q}'|= \sqrt{\hat{q}’^2}$, where we again make use of Eq.(37). Thus, the time component, $\sigma'$ of $q’$ will not appear in the wave functions of both P-wave and S-wave mesons in final state in the transition amplitude calculation. The BS normalizer of axial meson, $N_{A^-}$, can be obtained by solving the current conservation condition in Eq.(38).

We now obtain the $++$ and $--$ components of the axial meson wave function  with the corresponding adjoint wave functions as in case of P and V mesons done earlier, and calculate expressions for $++++$, $++--$, $--++$, and $----$ terms that appear in the scattering amplitude in Eq.(46).

We then evaluate trace over the gamma matrices in Eq.(46). We make use of the fact that $\hat{q}'=\hat{q}+\hat{m}_2 \hat{P}'$, where, $\hat{P}'=P'-\frac{P'.P}{P^2}P$. We combine various terms, and further make use of the fact that, for initial axial meson, $P.\epsilon^{\lambda}=0$. We can express $\hat{P}'.\epsilon= P'.\epsilon$. We can then express the invariant matrix element, $M^1_{fi}$ corresponding to Diagram 1 as,

\begin{equation}\label{9f3}
 M^1_{fi}=-ie N_A N_P\frac{1}{M} \int \frac{d^3 \hat q}{(2\pi)^3} \frac{\phi_{A}(\hat q')\phi_{P}(\hat q)}{16\omega_1\omega_2\omega'_1\omega'_2} [\Theta_1(\epsilon^{\lambda'}.\epsilon^{\lambda})
 +\Theta_2(\epsilon^{\lambda'}.P)(\epsilon^{\lambda}.P')+\Theta_3(\epsilon^{\lambda'}.P')(\epsilon^{\lambda}.P')],
\end{equation}

Similarly for Diagram 2, we write the invariant matrix element, $M^2_{fi}$ as,

\begin{equation}
M^2_{fi}=-ie N_A N_P\frac{1}{M} \int \frac{d^3 \hat q}{(2\pi)^3} \frac{\phi_{A}(\hat q')\phi_{P}(\hat q)}{16\omega_1\omega_2\omega'_1\omega'_2} [\Theta'_1(\epsilon^{\lambda'}.\epsilon^{\lambda})
 +\Theta'_2(\epsilon^{\lambda'}.P)(\epsilon^{\lambda}.P')+\Theta'_3(\epsilon^{\lambda'}.P')(\epsilon^{\lambda}.P')],
\end{equation}

Total amplitude $M_{fi}$ for the process can be expressed as the sum of amplitudes from the two diagrams, $M_{fi}=M^1_{fi}+M^2_{fi}$, where,

\begin{eqnarray}
&&\nonumber M_{fi}=R_1(\epsilon^{\lambda'}.\epsilon^{\lambda})+R_2(\epsilon^{\lambda'}.P)(\epsilon^{\lambda}.P')+R_3(\epsilon^{\lambda}.P')(\epsilon^{\lambda'}.P'),\\&&
\nonumber R_1=-ie N_A N_P\frac{1}{M^2} \int \frac{d^3 \hat q}{(2\pi)^3} \frac{\phi_{P}(\hat q')\phi_{A}(\hat q)}{16\omega_1\omega_2\omega'_1\omega'_2} (\Theta_1+\Theta'_1),\\&&
\nonumber R_2=-ieN_A N_P\frac{1}{M^2} \int \frac{d^3 \hat q}{(2\pi)^3} \frac{\phi_{P}(\hat q')\phi_{A}(\hat q)}{16\omega_1\omega_2\omega'_1\omega'_2}(\Theta_2+\Theta'_2),\\&&
 R_3=-ieN_A N_P\frac{1}{M^2} \int \frac{d^3 \hat q}{(2\pi)^3} \frac{\phi_{P}(\hat q')\phi_{A}(\hat q)}{16\omega_1\omega_2\omega'_1\omega'_2}(\Theta_3+\Theta'_3).
\end{eqnarray}

where integrals over $d^3\hat{q}$ are performed over $\Theta_i$, and $\Theta'_i$.

Thus, $R_1$, $R_2$, and $R_3$ are the three form factors in the above equation. Now, with a change of variables, and making use of the electromagnetic gauge invariance, it can be shown that these three form  factors appearing in $M_{fi}$, are not independent, and we can express the invariant amplitude, $M_{fi}$ in terms of a single form factor.

To show this, we introduce two new external variables, $\bar{P}$, and $k$, which are defined as,

\begin{equation}
\bar{P}=P + P'; k= P - P',
\end{equation}
where $\bar{P}$ is the sum of momenta of initial and emitted mesons, and $k$ is the emitted photon momentum. Thus, we can express the initial and final meson momenta $P$ and $P'$ in terms of new variables, as $P=\frac{\bar{P}+k}{2}$, and $P'=\frac{\bar{P}-k}{2}$. In terms of these new variables, we can express the amplitude, $M_{fi}$ as,

\begin{eqnarray}
&&\nonumber M_{fi}=S_1 (\epsilon'.\epsilon) +S_2 (\bar{P}.\epsilon)(\bar{P}.\epsilon')+S_3 (k.\epsilon)(\bar{P}.\epsilon')\\&&
S_1= R_1; S_2=\frac{1}{4}(R_2 +R_3); S_3 = -\frac{1}{4}(R_2 + R_3).
\end{eqnarray}
Now, the transversality property of polarization vector of axial meson, $P.\epsilon=0$ leads to,

\begin{equation}
\epsilon_{\nu}(\bar{P}+k)_{\nu}=0.
\end{equation}

We now introduce a new form factor, $S''=-S_2+S_3$, in terms of which we can write $M_{fi}$ as,

\begin{equation}
M_{fi}=\epsilon'_{\mu}\epsilon_{\nu} \bigg[S_1 \delta_{\mu,\nu}+ S_2\bar{P}_{\mu}(\bar{P}+ k)_{\nu}+ S''\bar{P}_{\mu}k_{\nu}\bigg]
\end{equation}

Now due to Eq.(55), the term with $S_2$ vanishes. Thus, amplitude, $M_{fi}$ can be expressed as,

\begin{eqnarray}
&&\nonumber M_{fi}=\epsilon'_{\mu}\epsilon_{\nu} M_{\mu \nu};\\&&
M_{\mu \nu}= S_1 \delta_{\mu \nu} +S'' \bar{P}_{\mu} k_{\nu}.
\end{eqnarray}

Now, electromagnetic gauge invariance demands, $k_{\mu} M_{\mu \nu}=0$. This leads to the relation,

\begin{equation}
S'' = -\frac{S_1}{\bar{P}.k},
\end{equation}
between the form factors, which is like an equation of constraint. Thus, due to the electromagnetic gauge invariance, $S_1$, and $S''$ are no longer independent, and we can express the amplitude, $M_{fi}$ in terms of a single form factor, $S_1$, whose expression is given in the next equations.

\begin{eqnarray}
&&\nonumber M_{fi} = S_1 \bigg[(\epsilon'.\epsilon)- \frac{1}{\bar{P}.k} (\bar{P}.\epsilon')(k.\epsilon)\bigg];\\&&
S_1=-ie N_A N_P\frac{1}{M^2} \int \frac{d^3 \hat q}{(2\pi)^3} \frac{\phi_{P}(\hat q')\phi_{A}(\hat q)}{16\omega_1\omega_2\omega'_1\omega'_2} (\Theta_1(\hat{q}^2)+\Theta'_1(\hat{q}^2))
\end{eqnarray}

The expressions for $\Theta_1(\hat{q}^2)$, and $\Theta'_1(\hat{q}^2)$ in the expression for $M_{fi}$ above are:

\begin{eqnarray}
&&\nonumber \Theta_1(\hat{q}^2)= \bigg[ (-\alpha_1 a_5 b_1- \alpha_2 a_5 d_1-\alpha_3 e_5 b_1-\alpha_4 e_5 d_1)(-4M^2)\hat{q}^2)+
 (-\alpha_1 a_8 b_1-\alpha_2 a_8 d_1- \alpha_3 e_8 b_1-\alpha_4 e_8 d_1) 4(P'.P) \hat{q}^2+\\&&
\nonumber (-\alpha_1 a_6 b_2 -\alpha_2 a_6 d_2 -\alpha_3 e_6 b_2 -\alpha_4 e_6 d_2)(-4M^2) (P'.P) \hat{q}^2+
 (-\alpha_1 a_7 b_2- \alpha_2 a_7 d_2 -\alpha_3 e_7 b_2- \alpha_4 e_7 d_2)(4M^2) \hat{q}^2+\\&&
\nonumber (\alpha_1 a_9 b_2 +\alpha_2 a_9 d_2 + \alpha_3 e_9 b_2 + \alpha_4 e_9 d_2)(-4M^2) (P'.P) \hat{q}^2+
(-\alpha_1 a_1 b_3 - \alpha_2 a_1 d_3- \alpha_3 e_1 b_3 - \alpha_4 e_1 d_3)(4M^2)\hat{q}^2+\\&&
\nonumber (-\alpha_1 a_4 b_3 -\alpha_2 a_4 d_3 - \alpha_3 e_4 b_3 -\alpha_4 e_4 d_3)(4M^2) (P'.P)\hat{q}^2+\\&&
\nonumber (-\alpha_1 a_8 b_3 -\alpha_2 a_8 d_3 -\alpha_3 e_8 b_3 -\alpha_4 e_8 d_3)(4M^2)[(P'.\hat{q}')\hat{q}^2+ (P'.\hat{q})\hat{q}^2]+\\&&
\nonumber (\alpha_1 a_2 b_4 +\alpha_2 a_2 d_4 +\alpha_3 e_2 b_4+\alpha_4 e_2 d_4)(4M^2)\hat{q}^2+
\nonumber (\alpha_1 a_3 b_4 +\alpha_2 a_3 d_4 +\alpha_3 e_3 b_4 +\alpha_4 e_3 d_4)(-4P'.P)\hat{q}^2+\\&&
\nonumber (\alpha_1 a_6 b_4 +\alpha_2 a_6 d_4 +\alpha_3 e_6 b_4 +\alpha_4 e_6 d_4)(4M^2 (\hat{q}'.P')\hat{q}^2+\\&&
(\alpha_1 a_9 b_4 +\alpha_2 a_9 d_4 +\alpha_3 e_9 b_4 +\alpha_4 e_9 d_4)[4M^2 \hat{q}^2 (P'.\hat{q}'+P'.\hat{q})]\bigg];
\end{eqnarray}

and
\bigskip
\begin{eqnarray}
&&\nonumber \Theta'_1(\hat{q}^2)=\bigg[(\alpha'_1 a_5 b_1+\alpha'_2 a_5 d_1 +\alpha'_3 e_5 b_1 +\alpha'_4 e_5 d_1)(-4M^2)\hat{q}^2+
(\alpha'_1 a_8 b_1 +\alpha'_2 a_8 d_1 +\alpha'_3 e_8 b_1 +\alpha'_4 e_8 d_1)4(P'.P)\hat{q}^2+\\&&
\nonumber (\alpha'_1 a_6 b_2+\alpha'_2 a_6 d_2 +\alpha'_3 e_6 b_2 +\alpha'_4 e_6 d_2)(-4M^2)(P'.P)\hat{q}^2+
(\alpha'_1 a_7 b_2 +\alpha'_2 a_7 d_2+ \alpha'_3 e_7 b_2 +\alpha'_4 e_7 d_2)(4M^2)\hat{q}^2+\\&&
\nonumber (\alpha'_1 a_9 b_2 +\alpha'_2 a_9 d_2+ \alpha'_3 e_9 b_2 +\alpha'_4 e_9 d_2)(-12M^2)(P'.P)\hat{q}^2+
(\alpha'_1 a_1 b_3 +\alpha'_2 a_1 d_3+ \alpha'_3 e_1 b_3 +\alpha'_4 e_1 d_3)(-4M^2)\hat{q}^2+\\&&
\nonumber (\alpha'_1 a_4 b_3 +\alpha'_2 a_4 d_3+ \alpha'_3 e_4 b_3 +\alpha'_4 e_4 d_3)(-4M^2)(P'.P)\hat{q}^2+
(\alpha'_1 a_8 b_3 +\alpha'_2 a_8 d_3+ \alpha'_3 e_8 b_3 +\alpha'_4 e_8 d_3)(-4M^2)(P'.\hat{q})\hat{q}^2+\\&&
\nonumber (\alpha'_1 a_2 b_4 +\alpha'_2 a_2 d_4+ \alpha'_3 e_2 b_4 +\alpha'_4 e_2 d_4)(-4M^2)\hat{q}^2+
(\alpha'_1 a_3 b_4 +\alpha'_2 a_3 d_4+ \alpha'_3 e_3 b_4 +\alpha'_4 e_3 d_4)4(P'.P)\hat{q}^2+\\&&
\nonumber (\alpha'_1 a_6 b_4 +\alpha'_2 a_6 d_4+ \alpha'_3 e_6 b_4 +\alpha'_4 e_6 d_4)(-4M^2)[(\hat{q'}.P')\hat{q}^2+(\hat{q}.P')\hat{q}^2]+\\&&
(\alpha'_1 a_9 b_4 +\alpha'_2 a_9 d_4+ \alpha'_3 e_9 b_4 +\alpha'_4 e_9 d_4)(4M^2)(P'.\hat{q}) \hat{q}^2\bigg],
\end{eqnarray}

which are expressible in terms of the dot products of momenta. Here, $\alpha_1,...,\alpha_4$, and $\alpha'_1,...,\alpha'_4$ are the results of contour integrals over the poles of the propagators in Eqs.(29), and (31) respectively, and the expressions for the coefficients, $a_i, b_i, d_i$ and $e_i$ entering into $\Theta_{1}$, and $\Theta'_{1}$ are:

\begin{eqnarray}
&&\nonumber a_1 =-\frac{M'}{4}-\frac{m_1 m_2 M'}{4\omega'_1\omega'_2}-\frac{\hat{q}^2M'}{4\omega'_1\omega'_2}; a_2=\frac{m_2M'}{4\omega'_2 M} +\frac{m_1 M'}{4\omega'_1 M}; a_3 = \frac{1}{4}+\frac{m_1 m_2}{4\omega'_1\omega'_2}+\frac{\hat{q'}^2}{4\omega'_1\omega'_2};\\&&
\nonumber a_4 = \frac{m_2}{4\omega'_2; M}+\frac{m_1}{4\omega'_1 M};
\nonumber a_5=\frac{M'}{4\omega'_2 M}+\frac{M'}{4\omega'_1 M};
a_6=\frac{-1}{4\omega'_2 M};
a_7 =\frac{M' m_1}{4\omega'_1\omega'_2}+\frac{M' m_2}{4\omega'_1\omega'_2} ; \\&&
\nonumber a_8 = \frac{m_1}{4\omega'_1 \omega'_2}+\frac{m_2}{4\omega'_1\omega'_2};
a_9=\frac{-1}{4\omega'_1 \omega'_2}; b_1=\frac{M}{4}+\frac{m_1 m_2 M}{4\omega_1 \omega_2}+ \frac{m_1 M}{4\omega_1}+ \frac{m_2 M}{4\omega_2}-\frac{\hat{q}^2 M}{4\omega_1 \omega_2}\\&&
\nonumber b_2=\frac{m_1}{4\omega_1}+\frac{m_2}{4\omega_2}+\frac{1}{4}+\frac{m_1 m_2}{4\omega_1 \omega_2}-\frac{\hat{q}^2}{4\omega_1 \omega_2}; b_3=-\frac{1}{4\omega_1}+\frac{1}{4\omega_2}-\frac{m_1}{4\omega_1 \omega_2}+\frac{m_2}{4\omega_1 \omega_2}\\&&
\nonumber b_4=-\frac{m_1 M}{4\omega_1 \omega_2}-\frac{M}{4\omega_2}+\frac{M}{4\omega_1}-\frac{m_2 M}{4\omega_1 \omega_2}; e_1=-\frac{M'}{4}-\frac{m_1 m_2 M'}{4\omega'_1 \omega'_2}+\frac{\hat{q'}^2 M'}{4\omega'_1\omega'_2}\\&&
\nonumber e_2=-\frac{m_2 M'}{4\omega'_2 M}-\frac{m_1 M'}{4\omega'_1 M}; e_3=\frac{1}{4}-\frac{m_1 m_2} {4\omega'_1\omega'_2}-\frac{\hat{q'}^2}{4\omega'_1\omega'_2}; e_4=-\frac{m_2}{4\omega'_2 M}+\frac{m_1}{4\omega'_1 M}\\&&
\nonumber e_5=\frac{M'}{4\omega'_2 M}- \frac{M'}{4\omega'_1 M}; e_6=-\frac{1}{4\omega'_2 M};  e_7=\frac{M' m_1}{4\omega'_1\omega'_2}+\frac{M' m_2}{4\omega'_1\omega'_2}\\&&
\nonumber e_8=-\frac{m_1}{4\omega'_1\omega'_2}-\frac{m_2}{4\omega'_1\omega'_2}; e_9=\frac{1}{4\omega'_1 M};
d_1=\frac{M}{4}+\frac{m_1 m_2 M}{4\omega_1\omega_2}-\frac{m_1 M}{4\omega_1}-\frac{m_2 M}{4\omega_2}-\frac{\hat{q}^2 M}{4\omega_1\omega_2};\\&&
\nonumber d_2= -\frac{m_1}{4\omega_1}-\frac{m_2}{4\omega_2+\frac{1}{4}}-\frac{m_1 m_2}{4\omega_1 \omega_2}-\frac{\hat{q}^2}{4\omega_1 \omega_2};
d_3=\frac{1}{4\omega_1}-\frac{1}{4\omega_2}-\frac{m_2}{4\omega_1 \omega_2}+\frac{m_1}{4\omega_1 \omega_2}\\&&
d_4=\frac{m_1 M}{4\omega_1\omega_2}+\frac{M}{4\omega_2}-\frac{M}{4\omega_1}+\frac{m_2 M}{4\omega_1\omega_2}
\end{eqnarray}
\bigskip

Now, to calculate the decay widths, we need to calculate the spin averaged amplitude square, $|\overline{M}_{fi}|^2$, where
$|\overline{M}_{fi}|^2=\frac{1}{2j+1}\sum_{\lambda,\lambda'}|{M}_{fi}|^2$, where we average over the initial polarization states $\lambda$ of A-meson, and sum over
the final polarization $\lambda'$ of photon. We make use of the normalizations,
$\Sigma_{\lambda} \epsilon_{\mu}^{\lambda}\epsilon_{\nu}^{\lambda}=\frac{1}{3}(\delta_{\mu\nu}+\frac{P_{\mu}P_{\nu}}{M^2})$ for vector meson, and
$\Sigma_{\lambda'} \epsilon_{\mu}^{\lambda'}\epsilon_{\nu}^{\lambda'}=\delta_{\mu\nu}$, for the emitted photon, with $M_{fi}$ taken from the previous
equation, we get, $\sum_{\lambda'}\sum_{\lambda}|\epsilon^{\lambda'}.\epsilon^{\lambda}|^2 =1$.

Making use of the kinematical relations,
\begin{eqnarray}
&&\nonumber \bar{P}.k=-M^2 + M'^2\\&&
\nonumber P.k= \frac{-M^2+M'^2}{2}\\&&
\nonumber \bar{P}^2=-2(M^2+ M'^2)\\&&
\bar{P}.P=\frac{-3M^2-M'^2}{2},
\end{eqnarray}

the spin-averaged amplitude square of the process can be written as
\begin{equation}
 |\overline{M}_{fi}|^2= \frac{1}{3}|S_1|^2 [1 - \frac{M'^2}{3M^2}].
\end{equation}

We can write the decay width,

\begin{equation}\label{50}
\Gamma_{A\rightarrow P\gamma}=\frac{|\overline{M}_{fi}|^2}{8\pi M^2}|\overrightarrow{P'}|,
\end{equation}
where we make use of the fact that modulus of the momentum of the emitted pseudoscalar meson can be expressed in terms of masses of particles as,
$|\overrightarrow{P'}|=\frac{1}{2M} (M^2-M'^2)$.

\section{Radiative decays of heavy-light quarkonia through $P\rightarrow A^-\gamma$}
We proceed to evaluate the process, $P\rightarrow A^-\gamma$ using Fig.\ref{fig:1}. Here, the initial pseudoscalar ($0^{-+}$) meson decays into an axial vector ($1^{+-}$) meson and a photon. We can then express the effective 3D form of invariant matrix element, $M^1_{fi}$ corresponding to Diagram 1 as,

\begin{multline}
 M^{1}_{fi}=-ie\int \frac{d^3 \hat q}{(2\pi)^3} \frac{1}{M^2}Tr\bigg[ \alpha_1 {\not}P\overline{\psi}_A^{++}(\hat q'){\not}\epsilon'\psi_P^{++}(\hat q)
 + \alpha_2 {\not}P\overline{\psi}_A^{++}(\hat q'){\not}\epsilon'\psi_P^{--}(\hat q)\\
 +\alpha_3 {\not}P\overline{\psi}_A^{--}(\hat q'){\not}\epsilon'\psi_P^{++}(\hat q)
 + \alpha_4 {\not}P\overline{\psi}_A^{--}(\hat q'){\not}\epsilon'\psi_P^{--}(\hat q)\bigg]
 \end{multline}

Similarly the amplitude for Diagram 2 can be written as,

\begin{multline}\label{7g}
 M^{2}_{fi}=-e_Q\int \frac{d^3 \hat q}{(2\pi)^3} \frac{1}{M^2}Tr\bigg[ \alpha'_1 \overline{\psi}_A^{++}(\hat q'){\not}P\psi_P^{++}(\hat q){\not}\epsilon'
 + \alpha'_2 \overline{\psi}_A^{++}(\hat q'){\not}P\psi_P^{--}(\hat q){\not}\epsilon'\\
 +\alpha'_3 \overline{\psi}_A^{--}(\hat q'){\not}P\psi_P^{++}(\hat q){\not}\epsilon'
 + \alpha'_4 \overline{\psi}_A^{--}(\hat q'){\not}P\psi_P^{--}(\hat q){\not}\epsilon'\bigg],
 \end{multline}

$\alpha_1,...,\alpha_4$, and $\alpha'_1,...,\alpha'_4$ in Eq.(29), and (31) are results of pole integrations over $Md\sigma$.

Taking trace over the gamma matrices in the previous equations, we get, the invariant matrix element, $M^1_{fi}$ for Diagram 1 as,

\begin{equation}\label{9f2}
 M^1_{fi}=-ie N_A N_P\frac{1}{M^2} \int \frac{d^3 \hat q}{(2\pi)^3} \frac{\phi_{A}(\hat q')\phi_{P}(\hat q)}{16\omega_1\omega_2\omega'_1\omega'_2} [\Delta_1(\epsilon^{\lambda'}.\epsilon^{\lambda})
 +\Delta_2(\epsilon^{\lambda'}.P)(\epsilon^{\lambda}.P)+\Delta_3(\epsilon^{\lambda'}.P')(\epsilon^{\lambda}.P)],
\end{equation}

Similarly for Diagram 2, we write the invariant matrix element, $M^2_{fi}$ as,

\begin{equation}
M^2_{fi}=-ie N_A N_P\frac{1}{M^2} \int \frac{d^3 \hat q}{(2\pi)^3} \frac{\phi_{A}(\hat q')\phi_{P}(\hat q)}{16\omega_1\omega_2\omega'_1\omega'_2} [\Delta'_1(\epsilon^{\lambda'}.\epsilon^{\lambda})
 +\Delta'_2(\epsilon^{\lambda'}.P)(\epsilon^{\lambda}.P)+\Delta'_3(\epsilon^{\lambda'}.P')(\epsilon^{\lambda}.P)],
\end{equation}

Total amplitude $M_{fi}$ for the process can be expressed as the sum of amplitudes from the two diagrams, $M_{fi}=M^1_{fi}+M^2_{fi}$, where,

\begin{eqnarray}
&&\nonumber M_{fi}=S'_1(\epsilon^{\lambda'}.\epsilon^{\lambda})+S'_2  (\epsilon^{\lambda}.P)(\epsilon^{\lambda'}.P)+S'_3(\epsilon^{\lambda}.P)(\epsilon^{\lambda'}.P')\\&&
\nonumber S'_1=-ie N_P N_A \int \frac{d^3 \hat q}{(2\pi)^3} \frac{\phi_{P}(\hat q)\phi_{A}(\hat q')}{16\omega_1\omega_2\omega'_1\omega'_2 M^4 M'}(\Delta_1(\hat{q}^2) + \Delta'_1(\hat{q}^2)) ,\\&&
\nonumber S'_2=-ie N_P N_A \int \frac{d^3 \hat q}{(2\pi)^3} \frac{\phi_{P}(\hat q)\phi_{A}(\hat q')}{16\omega_1\omega_2\omega'_1\omega'_2 M^4 M'} (\Delta_2(\hat{q}^2) + \Delta'_2(\hat{q}^2)),\\&&
\nonumber S'_3=-ie N_P N_A \int \frac{d^3 \hat q}{(2\pi)^3} \frac{\phi_{P}(\hat q)\phi_{A}(\hat q')}{16\omega_1\omega_2\omega'_1\omega'_2 M^4 M'}(\Delta_3(\hat{q}^2) + \Delta'_3(\hat{q}^2)) ,\\&&
\end{eqnarray}

The structure of $M_{fi}$ above is similar to \cite{wangjhep}.

To calculate the decay widths, we again need to calculate the spin averaged amplitude square, $|\overline{M}_{fi}|^2$, where
$|\overline{M}_{fi}|^2=\sum_{\lambda,\lambda'}|{M}_{fi}|^2$, where we sum over the final polarization states, $\lambda'$ of photon, and  $\lambda$ of V-meson. Following a similar procedure as in $A^- -> P \gamma$, it is seen that the contributions to spin averaged amplitude modulus square arises only from $S'_1$, while the contributions from $S'_2$, and $S'_3$ vanish after doing the averaging over the ploarization states.

And the spin averaged amplitude modulus square gives,
\begin{equation}
 |\overline{M}_{fi}|^2= |S'_1|^2,
\end{equation}
for evaluating which we need to evaluate the form factor $S_1$, for which we need to perform $\int \frac{d^3 \hat q}{(2\pi)^3}$ integration over $(\Delta_1(\hat{q}^2)) +\Delta'_1(\hat{q}^2))$. Expression for $\Delta_1(\hat{q}^2))$ that arises from Diagram 1 is,

\begin{eqnarray}
&&\nonumber \Delta_1(\hat{q}^2)= \bigg[4M^2\hat{q}^2(\alpha_1 b'_4 a'_1+\alpha_2 b'_4 d'_1 +\alpha_3 e'_4 a'_1+\alpha_4 e'_4 d'_1)+
4(P'.P)\hat{q}^2 (\alpha_1 b'_8 a'_1+\alpha_2 b'_8 d'_1 +\alpha_3 e'_8 a'_1+\alpha_4 e'_8 d'_1)+\\&&
\nonumber 4M^2 (P'.P)\hat{q}^2 (\alpha_1 b'_6 a'_2 +\alpha_2 b'_6 d'_2 +\alpha_3 e'_6 a'_2+\alpha_4 e'_6 d'_2)+\\&&
\nonumber 4M^2 \hat{q}^2 (\alpha_1 b'_7 a'_2 +\alpha_2 b'_7 d'_2 +\alpha_3 e'_7 a'_2+\alpha_4 e'_7 d'_2)+
-4M^2 (P'.\hat{q})\hat{q}^2 (\alpha_1 b'_6 a'_4 +\alpha_2 b'_6 d'_4 +\alpha_3 e'_6 a'_4+\alpha_4 e'_6 d'_4)+\\&&
\nonumber 4M^2 \hat{q}^2 (P.P') (\alpha_1 b'_9 a'_2 +\alpha_2 b'_9 d'_2 +\alpha_3 e'_9 a'_2+\alpha_4 e'_9 d'_2)+\\&&
\nonumber 4M^2 \hat{q}^2 (\alpha_1 b'_1 a'_3 +\alpha_2 b'_1 d'_3 +\alpha_3 e'_1 a'_3+\alpha_4 e'_1 d'_3)+
 4M^2 (P'.P) \hat{q}^2 (\alpha_1 b'_5 a'_3 +\alpha_2 b'_5 d'_3 +\alpha_3 e'_5 a'_3+\alpha_4 e'_5 d'_3)+\\&&
\nonumber 4M^2 (P'.\hat{q'})\hat{q}^2 (\alpha_1 b'_8 a'_3 +\alpha_2 b'_8 d'_3 +\alpha_3 e'_8 a'_3+\alpha_4 e'_8 d'_3)+
4M^2 \hat{q}^2 (\alpha_1 b'_2 a'_4 +\alpha_2 b'_2 d'_4 +\alpha_3 e'_2 a'_4+\alpha_4 e'_2 d'_4)-\\&&
4(P'.P)\hat{q}^2 (\alpha_1 b'_3 a'_4 +\alpha_2 b'_3 d'_4 +\alpha_3 e'_3 a'_4+\alpha_4 e'_4 d'_4)-
4M^2 (P'.\hat{q}')\hat{q}^2 (\alpha_1 b'_6 a'_4 +\alpha_2 b'_6 d'_4 +\alpha_3 e'_6 a'_4+\alpha_4 e'_6 d'_4)\bigg],
\end{eqnarray}

while, the expression for $\Delta'_1$ arising from for Diagram 2,

\begin{eqnarray}
&&\nonumber\Delta'_1(\hat{q}^2)= \bigg[-(\alpha'_1 b_6 a_4 +\alpha'_2 b'_6 d'_4 + \alpha'_3 e'_6 a'_4 +\alpha'_4 e'_6 d'_4)(P'.\hat{q}(4M^2)\hat{q}^2+
 (\alpha'_1 b'_4 a'_1 +\alpha'_2 b'_4 d'_1 + \alpha'_3 e'_4 a'_1 +\alpha'_4 e'_4 d'_1)(-4M^2)\hat{q}^2+\\&&
\nonumber (\alpha'_1 b'_8 a'_1 +\alpha'_2 b'_8 d'_1 + \alpha'_3 e'_8 a'_1 +\alpha'_4 e'_8 d'_1)4(P'.P)\hat{q}^2+
(\alpha'_1 b'_6 a'_2 +\alpha'_2 b'_6 d'_2 + \alpha'_3 e'_6 a'_2 +\alpha'_4 e'_6 d'_2) 8M^2(P'.P)\hat{q}^2+\\&&
\nonumber (\alpha'_1 b'_7 a'_2 +\alpha'_2 b'_7 d'_2 + \alpha'_3 e'_7 a'_2 +\alpha'_4 e'_7 d'_2) 4M^2 \hat{q}^2+
(\alpha'_1 b'_8 a'_3 +\alpha'_2 b'_8 d'_3 + \alpha'_3 e'_8 a'_3 +\alpha'_4 e'_8 d'_3)4M^2 (P'.\hat{q})\hat{q}^2+\\&&
\nonumber (\alpha'_1 b'_9 a'_2 +\alpha'_2 b'_9 d'_2 + \alpha'_3 e'_9 a'_2 +\alpha'_4 e'_9 d'_2)4M^2 (P'.P)\hat{q}^2+
(\alpha'_1 b'_1 a'_3 +\alpha'_2 b'_1 d'_3 + \alpha'_3 e'_1 a'_3 +\alpha'_4 e'_1 d'_3)4M^2 \hat{q}^2+\\&&
\nonumber (\alpha'_1 b'_5 a'_3 +\alpha'_2 b'_5 d'_3 + \alpha'_3 e'_5 a'_3 +\alpha'_4 e'_5 d'_3)4M^2 (P'.P)\hat{q}^2+
(\alpha'_1 b'_8 a'_3 +\alpha'_2 b'_8 d'_3 + \alpha'_3 e'_8 a'_3 +\alpha'_4 e'_8 d'_3)4M^2 (P'.\hat{q'})\hat{q}^2+\\&&
\nonumber (\alpha'_1 b'_2 a'_4 +\alpha'_2 b'_2 d'_4 + \alpha'_3 e'_2 a'_4 +\alpha'_4 e'_2 d'_4)4M^2 \hat{q}^2+
(\alpha'_1 b'_3 a'_4 +\alpha'_2 b'_3 d'_4 + \alpha'_3 e'_3 a'_4 +\alpha'_4 e'_3 d'_4) 4(P'.P)\hat{q}^2+\\&&
(\alpha'_1 b'_6 a'_4 +\alpha'_2 b'_6 d'_4 + \alpha'_3 e'_6 a'_4 +\alpha'_4 e'_6 d'_4)4M^2 (P'.\hat{q'}\hat{q}^2+ (\alpha'_1 b'_9 a'_4 +\alpha'_2 b'_9 d'_4 + \alpha'_3 e'_9 a'_4 +\alpha'_4 e'_9 d'_4)8M^2 (P'.\hat{q})\hat{q}^2\bigg],
\end{eqnarray}

which is again expressible in terms of dot products of various momenta, Again, $\alpha_1,...,\alpha_4$, and $\alpha'_1,...,\alpha'_4$ are the results of contour integrations over the poles of the propagators in Eqs.(29), and (31). The coefficients, $a_i, b_i, d_i, e_i$ entering into $\Delta_1$, and $\Delta'_1$ above are:

\begin{eqnarray}
&&\nonumber a'_1 = \frac{1}{4M^2 \sqrt{\omega_1\omega_2\omega'_1\omega'_2}}(M^3\omega_1\omega_2+m_1 \omega_2 M^3 + m_1 m_2 M^3+\omega_1 m_2 M^3-M^3 \hat{q}^2)\\&&
\nonumber a'_2=\frac{1}{4M^2 \sqrt{\omega_1\omega_2\omega'_1\omega'_2}}(M^2\omega_1\omega_2+m_1 \omega_2 M^2 + m_1 m_2 M^2+\omega_1 m_2 M^2- M^2 \hat{q}^2)\\&&
\nonumber a'_3 = \frac{1}{4M^2 \sqrt{\omega_1\omega_2\omega'_1\omega'_2}}(-M^2\omega_2 - M^2 m_2 + M^2 \omega_1 + M^2 m_1);
 a'_4 = \frac{1}{4M^2 \sqrt{\omega_1\omega_2\omega'_1\omega'_2}} (M^3 \omega_2 - M^3 m_2 - M^3 m_1 - M^3 \omega_1)\\&&
\nonumber b'_1 = \frac{1}{4M^2 \sqrt{\omega_1\omega_2\omega'_1\omega'_2}}(-M^2M' \omega'_1\omega'_2-m_1m_2 M' M^2 - M^2 M' \hat{q'}^2);
 b'_2 = \frac{1}{4M^2 \sqrt{\omega_1\omega_2\omega'_1\omega'_2}} (m_1M' M\omega'_2 +m_2M'M\omega'_1)\\&&
\nonumber b'_3= \frac{1}{4M^2 \sqrt{\omega_1\omega_2\omega'_1\omega'_2}} (M^2 \omega'_1\omega'_2 +M^2m_1m_2+M^2\hat{q'}^2); b'_4=\frac{1}{4M^2 \sqrt{\omega_1\omega_2\omega'_1\omega'_2}}(MM'\omega'_1+MM'\omega'_2)\\&&
\nonumber b'_5=\frac{1}{4M^2 \sqrt{\omega_1\omega_2\omega'_1\omega'_2}}(m_1 M\omega'_2 +m_2M\omega'_1);b'_6= \frac{1}{4M^2 \sqrt{\omega_1\omega_2\omega'_1\omega'_2}} (-M\omega'_1)\\&&
\nonumber b'_7 = \frac{1}{4M^2 \sqrt{\omega_1\omega_2\omega'_1\omega'_2}}(M^2 M' (m_2 + m_1);b'_8=\frac{1}{4M^2 \sqrt{\omega_1\omega_2\omega'_1\omega'_2}}M^2 (m_1 + m_2);b'_9 =\frac{1}{4M^2 \sqrt{\omega_1\omega_2\omega'_1\omega'_2}}(-M\omega'_2)\\&&
\nonumber e'_1=\frac{1}{4M^2 \sqrt{\omega_1\omega_2\omega'_1\omega'_2}}M^2(-M'\omega'_1\omega'_2-m_1m_2 M'+M'\hat{q}^2); e'_2=\frac{1}{4M^2 \sqrt{\omega_1\omega_2\omega'_1\omega'_2}}M' M (-m_1 \omega'_2- m_2 \omega'_1)\\&&
\nonumber e'_3=\frac{1}{4M^2 \sqrt{\omega_1\omega_2\omega'_1\omega'_2}} M^2 (\omega'_1 \omega'_2-m_1m_2 -\hat{q'}^2); d'_4 =\frac{1}{4M^2 \sqrt{\omega_1\omega_2\omega'_1\omega'_2}}M^3 (\omega_1 -\omega_2 + m_2 + m_1);\\&&
\nonumber e'_4=\frac{1}{4M^2 \sqrt{\omega_1\omega_2\omega'_1\omega'_2}}M M'(\omega'_1 -\omega'_2);e'_5=\frac{1}{4M^2 \sqrt{\omega_1\omega_2\omega'_1\omega'_2}}M (m_1 \omega'_2 - m_2\omega'_1);e'_6=\frac{1}{4M^2 \sqrt{\omega_1\omega_2\omega'_1\omega'_2}}(-M\omega'_1)\\&&
\nonumber e'_7= \frac{1}{4M^2 \sqrt{\omega_1\omega_2\omega'_1\omega'_2}}M' M^2 (m_1+m_2);e'_8=\frac{1}{4M^2 \sqrt{\omega_1\omega_2\omega'_1\omega'_2}}M^2 (-m_1-m_2);e'_9=\frac{1}{4M^2 \sqrt{\omega_1\omega_2\omega'_1\omega'_2}}(M\omega'_2)\\&&
\nonumber d'_1=\frac{1}{4M^2 \sqrt{\omega_1\omega_2\omega'_1\omega'_2}}M^3(\omega_1\omega_2-m_1\omega_2+ m_1 m_2-\omega_1 m_2 -\hat{q}^2); d'_3=\frac{1}{4M^2 \sqrt{\omega_1\omega_2\omega'_1\omega'_2}}M^2 (\omega_2 -\omega_1+ m_1 -m_2)\\&&
d'_2=\frac{1}{4M^2 \sqrt{\omega_1\omega_2\omega'_1\omega'_2}}M^2(-m_1\omega_2+ \omega_1\omega_2- \omega_1 m_2- m_1 m_2- \hat{q}^2).
\end{eqnarray}

The decay widths $\Gamma$ for the process, $P \rightarrow A^- \gamma$, are given by Eq.(\ref{50}), with $P'$, now the momentum of the emitted axial meson.

\begin{table}
\begin{center}
\begin{tabular}{p{3cm} p{1.7cm} p{2.7cm} p{3.1cm} p{2.1cm} p{2.3cm} p{1.7cm} }
  \hline\hline
        &$\beta$ & BSE-CIA &  Expt.  & \cite{ebert03}    & PM  &RQM   \\
   \hline
 $\Gamma_{h_c(1P)\rightarrow \eta_c (1S_0)\gamma}$&0.253& 363.047 & 357$\pm$204\cite{zyla2020}& 560\cite{ebert03}& 398$\pm$99\cite{shi17}& 482 \cite{brambila04}\\
$\Gamma_{h_c(2P)\rightarrow \eta_c (2S_0)\gamma}$&0.471&187.145& & &160\cite{deng17} &218\cite{godfrey05}\\
$\Gamma_{h_c(2P)\rightarrow \eta_c (1S_0)\gamma}$&0.510&20.195 & & &135\cite{deng17} &85\cite{godfrey05} \\
 $\Gamma_{\eta_c(2S_0)\rightarrow h_c(1P)\gamma}$&0.650& 6.909 &  &6.2 \cite{ebert03}  &49\cite{godfrey05} & \\
   \hline\hline
  \end{tabular}
\caption{Radiative decay widths of heavy-light mesons (in Kev) for E1 transitions in BSE, along with experimental data  and results of other models.}
\end{center}
\end{table}

\section{Results and Discussion}
The present work is an extension of our work in \cite{bhatnagar20} to study of radiative M1 decays,  $P\rightarrow V \gamma$, and E1 decays, $A^- \rightarrow P\gamma$, and $P\rightarrow A^- \gamma$ of heavy-light quarkonia in the framework of $4\times 4$ BSE under Covariant Instantaneous Ansatz (CIA), which is a Lorentz-invariant generalization of Instantaneous Approximation. In our recent work \cite{bhatnagar20}, we had studied the processes, $V\rightarrow P \gamma$, $V\rightarrow S \gamma$, and $S-> V\gamma$. Such processes involve quark-triangle diagrams, and involve two hardon-quark vertices and are difficult to evaluate in BSE under CIA. We have made use of the generalized method of handling quark triangle diagrams with two hadron-quark vertices  in the framework of $4\times 4$ BSE, by expressing the transition amplitude, $M_{fi}$ as a linear superposition of terms (shown in \cite{bhatnagar20}) involving all possible combinations of $++$, and $--$ components of Salpeter wave functions of final and initial hadrons, through the terms, $++++$, $----$, $++--$, and $--++$, with each of the four terms being associated with a coefficient, $\alpha_i (i=1,...,4)$, which is the result of pole integration in the complex $\sigma$-plane. This superposition of all possible terms is a feature of relativistic frameworks.

In our previous work\cite{bhatnagar20}, we had simplified the calculation by considering only the most leading Dirac structures in the wave functions of P, V and A mesons, that contribute maximum to calculation of all meson observables in accordance with our power counting rule\cite{bhatnagar06,bhatnagar14}.  However in the present calculation, we consider the two leading order Dirac structures in the BS wave functions of P, and A mesons given in Eqs.(35), and (48) in accordance with the power counting scheme we proposed in \cite{bhatnagar06,bhatnagar14}, which makes this calculation more rigorous.

Using this generalized expression for $M_{fi}$, in Eq.(28-31), we have evaluated the decay widths for $M1$ transitions, $^1S_0\rightarrow ^3S_1 +\gamma$, involving the decays of the ground and excited states of the heavy-light mesons such as, $\eta_c(2S), B_c(2S)$. We wish to mention that as seen from Tables 2 and 3, the decay rates of M1 transitions are much weaker than the rates for E1 transitions. But M1 decay rates are interesting as they allow access to spin-singlet states, that are very difficult to produce. However, the known M1 decay rates show a serious disagreement between theory and experiment, as can be seen from Table 2.

As regards the $E1$ transitions, we have studied the processes, $^1P_1 \rightarrow ^1S_0 +\gamma$, that involve the decays, $h_c(1P)-> \eta_c(1S)\gamma$, $h_c(2P)->\eta_c(2S)\gamma$, and $h_c(2P)->\eta_c(1S)\gamma$, and the processes, $^1S_0 \rightarrow ^1P_1 +\gamma$, that involve the decays, $\eta_c(2S) \rightarrow h_c(1P)\gamma$.

We used algebraic forms of 3D Salpeter wave functions obtained through analytic solutions of mass spectral equations in approximate harmonic oscillator basis for ground and excited states of $0^{-+},1^{--}$, and $1^{+-}$ heavy-light quarkonia for calculation of their decay widths. The input parameters used by us are: $C_0$= 0.69, $\omega_0$= 0.22 GeV, $\Lambda_{QCD}$= 0.25 GeV, and $A_0$= 0.01, along with the input quark masses $m_u$= 0.30 GeV, $m_s$= 0.43 GeV, $m_c$= 1.49 GeV, and $m_b$= 4.67 GeV., that were obtained by fitting to their mass spectra\cite{eshete19}. We have compared our results with experimental data, where ever available, and other models, and found reasonable agreements.

Similarly we again see a wide range of variations in different models for both $M1$, and $E1$ transitions, particularly for decays of $\eta_c$, and $h_c$ mesons. Further, our decay widths for $nS -> n'S$ transitions in M1 decays, and $nP -> n'S$ transitions in E1 decays show a marked decrease as we go from ground to higher excited states, which is in conformity with data and other models. We have also given our predictions for radiative decays, $h_c(2P)->\eta_c(2S)\gamma$, $h_c(2P)->\eta_c(1S)\gamma$, and $\eta_c(2S)\rightarrow h_c(1P)$ for which data is not yet available. As regards M1 transitions, we have given our prediction for the decay width of $\eta_c(2S)\rightarrow J/\Psi(1S)\gamma$, for which the PDG tables\cite{zyla2020} give only the upper limit on the decay width. Also we calculated the decay width $B_c(2S)\rightarrow B^*_c(1S)\gamma$ for which data is not available.

The aim of doing this work was mainly to study the processes, $P\rightarrow V\gamma$, $A^-\rightarrow P\gamma$, and $P\rightarrow A^-\gamma$ for which very little data is available. This study was also to test the algebraic forms of wave functions of $A^-$ mesons\cite{vaishali21} that we have recently derived from the mass spectral equations of these axial mesons, along with the wave functions of P mesons and V mesons by studying their transitions. These wave functions were obtained as solutions of their mass spectral equations in an approximate harmonic oscillator basis obtained analytically from $4 \times 4$ BSE as a starting point, that has so far given good predictions \cite{eshete19,bhatnagar18, hluf16} not only of the mass spectrum of heavy-light quarkonia, but also their leptonic decays, two-photon, and two gluon decays. The present work would in turn lead to the validation of our approach, which provides a much deeper insight than the purely numerical calculations in $4 \times 4$ BSE approach that are prevalent in the literature.

A more detailed study on the transition form factors of both $M1$, and $E1$ transitions, but also the "static" form factors describing meson-photon interactions through the vertex $M\gamma M$ for various mesons will be relegated to a separate paper.

\bigskip

   \appendix
\section{Appendix}
\subsection{Radiative decays through $P \rightarrow V \gamma$}
Substituting the 3D BS wave function of pseudoscalar meson in Eq.(\ref{wf44}), we obtain the $++$ and $--$ components as
\begin{align}\label{d5}
\nonumber  \psi_P^{++}(\hat q)=\frac{N_P\phi_P(\hat q)}{4\omega_1\omega_2}[M((\omega_1\omega_2+m_1m_2+\hat q^2)+(m_1\omega_2+\omega_1m_2))-i((m_1\omega_2+\omega_1m_2)+(\omega_1\omega_2+m_1m_2-\hat q^2)){\not}P\\ \nonumber
  +iM((\omega_1-\omega_2)+(m_1-m_2)){\not}\hat q +((\omega_1+\omega_2)+(m_1+m_2)){\not}P{\not}\hat q]\gamma_5\\\nonumber
    \psi_P^{--}(\hat q)=\frac{N_P\phi_P(\hat q)}{4\omega_1\omega_2}[M(-(\omega_1\omega_2+m_1m_2+\hat q^2)+(m_1\omega_2+\omega_1m_2))-i((m_1\omega_2+\omega_1m_2)+(\omega_1\omega_2+m_1m_2-\hat q^2)){\not}P\\
  -iM((\omega_1-\omega_2)-(m_1-m_2)){\not}\hat q +(-(\omega_1+\omega_2)+(m_1+m_2)){\not}P{\not}\hat q]\gamma_5
  \end{align}

The adjoint Bethe-Salpeter wave function of pseudoscalar meson can be obtained by evaluating
$\overline{\psi}^{\pm\pm}_P(\hat q')=\gamma_4(\psi^{\pm\pm}_P(\hat q'))^+\gamma_4$ as
\begin{align}\label{d6}
\nonumber  \overline{\psi}_P^{++}(\hat q)=\frac{N_P\phi_P(\hat q)}{4\omega_1\omega_2}[-M((\omega_1\omega_2+m_1m_2+\hat q^2)+(m_1\omega_2+\omega_1m_2))+i((m_1\omega_2+\omega_1m_2)+(\omega_1\omega_2+m_1m_2-\hat q^2)){\not}P\\ \nonumber
  -iM((\omega_1-\omega_2)+(m_1-m_2)){\not}\hat q -((\omega_1+\omega_2)+(m_1+m_2)){\not}P{\not}\hat q]\gamma_5\\\nonumber
    \overline{\psi}_P^{--}(\hat q)=\frac{N_P\phi_P(\hat q)}{4\omega_1\omega_2}[-M(-(\omega_1\omega_2+m_1m_2+\hat q^2)+(m_1\omega_2+\omega_1m_2))-i((m_1\omega_2+\omega_1m_2)+(\omega_1\omega_2+m_1m_2-\hat q^2)){\not}P\\
  +iM((\omega_1-\omega_2)-(m_1-m_2)){\not}\hat q -(-(\omega_1+\omega_2)+(m_1+m_2)){\not}P{\not}\hat q]\gamma_5
\end{align}

Following the same steps as in Eq.(\ref{d71}), we obtain the $++$ and $--$ components of vector meson wave function in Eq.(\ref{wf44}) as
\begin{align}\label{d71}
\nonumber \psi_V^{++}(\hat q')= \frac{N_V\phi_V(\hat q)}{4\omega'_1\omega'_2} [ iM'\omega'_1\omega'_2{\not}\epsilon
  -\frac{M'}{M}\omega'_1m_2 {\not}\epsilon{\not}P +\frac{iM'}{M}\omega'_1{\not}\epsilon{\not}P{\not}\hat q'
  +\omega'_1\omega'_2 {\not}\epsilon{\not}P'
  +  \frac{im_2\omega'_1}{M} {\not}\epsilon{\not}P'{\not}P+\frac{\omega'_1}{M}{\not}\epsilon{\not}P'{\not}P{\not}\hat q' +\frac{M'}{M}\omega_2'm_1{\not}P{\not}\epsilon\\ \nonumber
  +\frac{iM'}{M^2}m_1m_2 {\not}P{\not}\epsilon{\not}P+\frac{M'}{M^2}m_1{\not}P{\not}\epsilon{\not}P{\not}\hat q'-\frac{im_1\omega'_2}{M}{\not}P{\not}\epsilon{\not}P'+\frac{m_1m_2}{M^2}{\not}P{\not}\epsilon{\not}P'{\not}P-\frac{im_1}{M^2} {\not}P{\not}\epsilon{\not}P'{\not}P{\not}\hat q'-\frac{iM'}{M}\omega'_2  \nonumber{\not}\hat q'{\not}P{\not}\epsilon\\
  +\frac{M'}{M^2}m_2{\not}\hat q'{\not}P{\not}\epsilon{\not}P-\frac{iM'}{M^2} {\not}\hat q'{\not}P{\not}\epsilon{\not}P{\not}\hat q'-\frac{\omega'_2}{M}{\not}\hat q'{\not}P{\not}\epsilon{\not}P'-\frac{im_2}{M^2}{\not}\hat q'{\not}P{\not}\epsilon{\not}P'{\not}P-\frac{1}{M^2}{\not}\hat q'{\not}P{\not}\epsilon{\not}P'{\not}P{\not}\hat q'\\ \nonumber
  \psi_V^{--}(\hat q')= \frac{N_V\phi_V(\hat q)}{4\omega'_1\omega'_2} [ iM'\omega'_1\omega'_2{\not}\epsilon
  +\frac{M'}{M}\omega'_1m_2 {\not}\epsilon{\not}P -\frac{iM'}{M}\omega'_1{\not}\epsilon{\not}P{\not}\hat q'
  +\omega'_1\omega'_2 {\not}\epsilon{\not}P'
  -  \frac{im_2\omega'_1}{M} {\not}\epsilon{\not}P'{\not}P-  \frac{\omega'_1}{M}{\not}\epsilon{\not}P'{\not}P{\not}\hat q' -\frac{M'}{M}\omega_2'm_1{\not}P{\not}\epsilon\\ \nonumber
  +\frac{iM'}{M^2}m_1m_2 {\not}P{\not}\epsilon{\not}P+\frac{M'}{M^2}m_1{\not}P{\not}\epsilon{\not}P{\not}\hat q'+\frac{im_1\omega'_2}{M}{\not}P{\not}\epsilon{\not}P'+\frac{m_1m_2}{M^2}{\not}P{\not}\epsilon{\not}P'{\not}P-\frac{im_1}{M^2} {\not}P{\not}\epsilon{\not}P'{\not}P{\not}\hat q'+\frac{iM'}{M}\omega'_2  \nonumber{\not}\hat q'{\not}P{\not}\epsilon\\
  +\frac{M'}{M^2}m_2{\not}\hat q'{\not}P{\not}\epsilon{\not}P-\frac{iM'}{M^2} {\not}\hat q'{\not}P{\not}\epsilon{\not}P{\not}\hat q'+\frac{\omega'_2}{M}{\not}\hat q'{\not}P{\not}\epsilon{\not}P'-\frac{im_2}{M^2}{\not}\hat q'{\not}P{\not}\epsilon{\not}P'{\not}P-\frac{1}{M^2}{\not}\hat q'{\not}P{\not}\epsilon{\not}P'{\not}P{\not}\hat q'\\ \nonumber,
\end{align}
where as the adjoint wave functions are
\begin{align}\label{d72}
\nonumber \overline{\psi}_V^{++}(\hat q')= \frac{N_V\phi_V(\hat q)}{4\omega'_1\omega'_2} [ iM'\omega'_1\omega'_2{\not}\epsilon
  -\frac{M'}{M}\omega'_1m_2 {\not}P{\not}\epsilon +\frac{iM'}{M}\omega'_1{\not}\hat q'{\not}P{\not}\epsilon
  +\omega'_1\omega'_2 {\not}P'{\not}\epsilon
  +  \frac{im_2\omega'_1}{M} {\not}P{\not}P'{\not}\epsilon+\frac{\omega'_1}{M}{\not}\hat q'{\not}P{\not}P'{\not}\epsilon +\frac{M'}{M}\omega_2'm_1{\not}\epsilon{\not}P\\ \nonumber
  +\frac{iM'}{M^2}m_1m_2 {\not}P{\not}\epsilon{\not}P+\frac{M'}{M^2}m_1{\not}\hat q'{\not}P{\not}\epsilon{\not}P-\frac{im_1\omega'_2}{M}{\not}P'{\not}\epsilon{\not}P+\frac{m_1m_2}{M^2}{\not}P{\not}P'{\not}\epsilon{\not}P-\frac{im_1}{M^2} {\not}\hat q'{\not}P{\not}P'{\not}\epsilon{\not}P-\frac{iM'}{M}\omega'_2 {\not}\epsilon{\not}P{\not}\hat q'\\ \nonumber
  +\frac{M'}{M^2}m_2{\not}P{\not}\epsilon{\not}P{\not}\hat q'-\frac{iM'}{M^2} {\not}\hat q'{\not}P{\not}\epsilon{\not}P{\not}\hat q'-\frac{\omega'_2}{M}{\not}P'{\not}\epsilon{\not}P{\not}\hat q'-\frac{im_2}{M^2}{\not}P{\not}P'{\not}\epsilon{\not}P{\not}\hat q'-\frac{1}{M^2}{\not}\hat q'{\not}P{\not}P'{\not}\epsilon{\not}P{\not}\hat q'\\ \nonumber
  \overline{\psi}_V^{--}(\hat q')= \frac{N_V\phi_V(\hat q)}{4\omega'_1\omega'_2} [ iM'\omega'_1\omega'_2{\not}\epsilon
  +\frac{M'}{M}\omega'_1m_2 {\not}P{\not}\epsilon -\frac{iM'}{M}\omega'_1{\not}\hat q'{\not}P{\not}\epsilon
  +\omega'_1\omega'_2 {\not}P'{\not}\epsilon
  - \frac{im_2\omega'_1}{M} {\not}P{\not}P'{\not}\epsilon-\frac{\omega'_1}{M}{\not}\hat q'{\not}P{\not}P'{\not}\epsilon -\frac{M'}{M}\omega_2'm_1{\not}\epsilon{\not}P\\  \nonumber
  +\frac{iM'}{M^2}m_1m_2 {\not}P{\not}\epsilon{\not}P+\frac{M'}{M^2}m_1{\not}\hat q'{\not}P{\not}\epsilon{\not}P+\frac{im_1\omega'_2}{M}{\not}P'{\not}\epsilon{\not}P+\frac{m_1m_2}{M^2}{\not}P{\not}P'{\not}\epsilon{\not}P-\frac{im_1}{M^2} {\not}\hat q'{\not}P{\not}P'{\not}\epsilon{\not}P+\frac{iM'}{M}\omega'_2 {\not}\epsilon{\not}P{\not}\hat q'\\
  +\frac{M'}{M^2}m_2{\not}P{\not}\epsilon{\not}P{\not}\hat q'-\frac{iM'}{M^2} {\not}\hat q'{\not}P{\not}\epsilon{\not}P{\not}\hat q'+\frac{\omega'_2}{M}{\not}P'{\not}\epsilon{\not}P{\not}\hat q'-\frac{im_2}{M^2}{\not}P{\not}P'{\not}\epsilon{\not}P{\not}\hat q'-\frac{1}{M^2}{\not}\hat q'{\not}P{\not}P'{\not}\epsilon{\not}P{\not}\hat q'
\end{align}

Using above expressions, We calculate ${\not}P\overline{\psi}^{++}_{P}(\hat q'){\not}\epsilon'\Psi^{++}_{V}(\hat q)$,  ${\not}P\overline{\psi}^{++}_{P}(\hat q'){\not}\epsilon'\Psi^{--}_{V}(\hat q)$, ${\not}P\overline{\psi}^{--}_{P}(\hat q'){\not}\epsilon'\Psi^{++}_{V}(\hat q)$, and ${\not}P\overline{\psi}^{--}_{P}(\hat q'){\not}\epsilon'\Psi^{--}_{V}(\hat q)$, which are employed in the calculation of transition form factor, $F_{PV}$ for $P\rightarrow V\gamma$. The contribution of $F_{PV}$ from Diagram 1 is given by,

\begin{eqnarray}
&&\nonumber F^1_{PV}= e_q N_V N_P\frac{1}{M^2} \int\frac{d^3\hat q}{(2\pi)^3}\frac{\phi_V(\hat q')\phi_P(\hat q)}{16\omega_1\omega_2\omega'_1\omega'_2} [TR_1];\\&&
\nonumber [TR_1]=(\alpha_1a_4+\alpha_2b_4+\alpha_3a_4+\alpha_4b_4)\bigg(- M'\omega'_1 \omega'_2 \frac{M'^2-M^2}{2MM'^2}|\hat{q}|+
 M'm_1m_2\frac{(M^2-M'^2)|\hat{q}|}{2M M'^2}
+\\&&
\nonumber
m_1 (\hat{q}^2- \hat{m}_2 \frac{M^2-M'^2}{2M}|\hat{q}|)
 +
 M'(\frac{2}{3}\hat{m}_2\hat{q}^2
- \frac{\hat{q}^2 (M^2-M'^2)}{2M M'^2}|\hat{q}|)
 - m_2 (\hat{q}^2 +\hat{m}_2\frac{M^2-M'^2}{2M}|\hat{q}|)\bigg)\\&& \nonumber
 +(\alpha_1a_3+\alpha_2b_3-\alpha_3a_3-\alpha_4b_3)\bigg(M M' m_2\omega'_1\frac{M^2-M'^2}{M'^2(2M)}|\hat{q}|
+M\omega'_1(-\hat{q}^2)
+
 M'\omega'_2 m_1\frac{M^2-M'^2}{2M'^2}|\hat{q}| -\omega'_2\hat q^2\bigg)
\\&&
\nonumber +(\alpha_1a_1+\alpha_2b_1+\alpha_3a_1+\alpha_4b_1)  \bigg( \omega'_1\omega'_2
+
M' m_1 (\hat{m}_2 -\frac{(M^2-M'^2)|\hat{q}|}{2M M'^2})
 -  m_1 m_2
-
 M' m_2 (\hat{m}_2 -\frac{M^2-M'^2}{2M M'^2}|\hat{q}|)
\\&&
\nonumber -(-3\hat{q}^2+ \hat{m}_2 \frac{M^2-M'^2}{2M}|\hat{q}|+\frac{(M^2-M'^2)}{2M}|\hat{q}|+\frac{(M^2-M'^2)^2}{4M^2})  \bigg)
\\&&
\nonumber  +(\alpha_1a_3+\alpha_2b_3+\alpha_3a_3+\alpha_4b_3) \bigg(M' m_1 \hat{m}_2 \frac{(M^4-M'^4)|\hat{q}}{8 M M'^2}
 -  \frac{1}{2}M'm_2 \frac{\hat{m}_2(M^4-M'^4)}{2M M'}\\&& \nonumber +(-\frac{2}{3}\hat m_2 (\hat q^2-\hat m_2^2\frac{(M^2-M'^2)^2}{4M^2})\frac{(M^2+M'^2)}{2}-  2\hat m_2^2\frac{1}{M'^2}(M'^2-\frac{(M^2+M'^2)^2}{4M^2})^2\hat m_2 \frac{(M^2+M'^2)}{2})\bigg)\\&& \nonumber
+(\alpha_1a_2+\alpha_2b_2-\alpha_3a_2-\alpha_4b_2) \bigg(-MM'\omega'_1 (-\hat{m}_2+ \frac{M^2-M'^2}{2M M'^2}|\hat{q}|)+ M m_2\omega'_1+m_1 \omega'_2 M \\&& \nonumber
 +\frac{M'}{M}\omega'_2 (M^2- \frac{M}{2M'^2}(M^2-M'^2)|\hat{q}|)
 \bigg)
-(\alpha_1a_2+\alpha_2b_2+\alpha_3a_2+\alpha_4b_2) m_1 \hat{m}_2\frac{M^2+M'^2}{2}\\&&
\nonumber +(-\alpha_1a_4-\alpha_2b_4+\alpha_3a_4+\alpha_4b_4) \bigg(M M'\omega'_1\frac{(M^4- M'^4)}{4M^3 M'^2}|\hat{q}|-\frac{M'}{M}\omega'_2\frac{(M^4-M'^4)}{4MM'^2}|\hat q|\bigg)\\&&
-(\alpha_1a_1+\alpha_2b_1-\alpha_3a_1-\alpha_4b_1)\bigg(M\omega'_1\hat{m}_2\frac{M^2+M'^2}{2M^2}+\frac{\omega'_2}{M}\hat m_2\frac{(M^2+M'^2)}{2}\bigg)
\end{eqnarray}

The contribution to transition form factor $F_{PV}$ from Diagram 2 is,

\begin{eqnarray}
&&\nonumber F^2_{PV}= e_QN_VN_P \frac{1}{M^2}\int\frac{d^3\hat q}{(2\pi)^3}\frac{\phi_V(\hat q')\phi_P(\hat q)}{16\omega_1\omega_2\omega'_1\omega'_2}[TR_2];\\&&
\nonumber [TR_2]= (-\alpha'_1a_3-\alpha'_2b_3+\alpha'_3a_3+\alpha'_4b_3)\bigg(-\frac{(M^2-M'^2)}{2MM'^2}|\hat q| -M\omega'_1(-\hat q^2+\hat m_2\frac{(M^2-M'^2)}{2M}|\hat q|)\\&& \nonumber +MM'\omega'_2m_1(\frac{(M^2-M'^2)}{2MM'^2}|\hat q|) +M\omega'_2(\hat q^2+\frac{5}{3}\hat m_2\frac{(M^2-M'^2)}{2M}|\hat q|) \bigg)\\&&
\nonumber +(\alpha'_1a_2+\alpha'_2b_2-\alpha'_3a_2-\alpha'_4b_2)\bigg(MM'\omega'_1(\hat m_2+\frac{(M^2-M'^2)}{2MM'^2}|\hat q|)   +Mm_2\omega'_1-Mm_1\omega'_2\\&&
\nonumber -MM'\omega'_2(\hat m_2-\frac{(M^2-M'^2)}{2MM'^2}|\hat q|)-m_2M^2] \bigg)\\&&
\nonumber +(-\alpha'_1a_4-\alpha'_2b_4+\alpha'_3a_4+\alpha'_4b_4)\bigg(\frac{1}{2}\hat m_2\frac{(M^4-M'^4)}{2MM'^2}|\hat q|  -\frac{m_1}{M^2}(\frac{1}{3}M^2\hat q^2+M^2\hat m_2^2\frac{(M^2-M'^2)}{2M}|\hat q|\\&&
\nonumber -\frac{M'}{M}\omega'_2\hat m_2\frac{(M^4-M'^4)}{4MM'^2}|\hat q| ) \bigg) \\&&
\nonumber +(\alpha'_1a_1+\alpha'_2b_1+\alpha'_3a_1+\alpha'_4b_1) \bigg(\omega'_1\omega'_2+\frac{M'}{M^2}m_1 (-\frac{M^2}{M^2}\frac{(M^2-M'^2)}{2M}|\hat q|+\hat m_2M^2)- m_1m_2\\&&
\nonumber + \frac{M'}{M^2}m_2(\frac{M}{M'^2}\frac{(M^2-M'^2)}{2}|\hat q|-\hat m_2M^2)+ (\hat q^2+2\hat m_2\frac{(M^2-M'^2)}{2M}|\hat q|+ \hat m_2^2\frac{(M^2-M'^2)^2}{4M^2})  \bigg)\\&&
\nonumber +(\alpha'_1a_3+\alpha'_2b_3+\alpha'_3a_3+\alpha'_4b_3) \bigg(-M'm_1\hat m_2\frac{(M^4-M'^4)}{2MM'^2}|\hat q|+ M^2m_2(\hat m_2\frac{M}{M'^2}\frac{(M^2-M'^2)}{2}|\hat q|\\&&
\nonumber -\frac{\hat m_2}{M'^2}\frac{(M^4-M'^4)}{4M}|\hat q|) +(M^2+M'^2)(\hat q^2\frac{(M^2-M'^2)}{2MM'^2}|\hat q|-2\hat m_2\hat q^2)-M^2(\frac{1}{2}\hat m_2\hat q^2\frac{(M^2+M'^2)}{M^2}\\&&
\nonumber +\hat m_2(\frac{\hat q^2}{M^2}(M^2+M'^2)+\frac{(M^2+M'^2)}{2M^2}\hat q^2)+ \hat m_2^2(M'^2-\frac{(M^2+M'^2)^2}{M^2}\frac{(M^2-M'^2)}{2M}|\hat q|))\\&&
\nonumber -(M^2+M'^2)\hat m_2^2(-M'^2+\frac{(M^2+M'^2)^2}{M^2})\frac{(M^2-M'^2)|\hat q|}{2MM'^2}    \bigg)
\nonumber +(\alpha'_1a_1+\alpha'_2b_1-\alpha'_3a_1-\alpha'_4b_1)\bigg(-\frac{\omega'_2}{M}(M^2+M'^2)\hat m_2 \bigg)\\&&
\nonumber +(\alpha'_1a_4+\alpha'_2b_4-\alpha'_3a_4-\alpha'_4b_4)\bigg(-\frac{M'}{M^2}m_1m_2\frac{M^2}{M'^2}\frac{(M^2-M'^2)}{2M}|\hat q|+ \frac{M'}{M^2}(\frac{M^2}{M'^2}\hat q^2\frac{(M^2-M'^2)}{2M}|\hat q|\\&&  -\frac{2}{3}M^2\hat m_2^2\frac{(M^2-M'^2)}{2M} |\hat q|- m_2M'^2\frac{(M^2-M'^2)}{2M}|\hat q|-\hat m_2^2\frac{(M^2+M'^2)}{4M'^2}\frac{(M^2-M'^2)}{2M}|\hat q|-m_2\hat q^2 \bigg)
\end{eqnarray}

where,
\begin{eqnarray}
&&\nonumber a_1= M((\omega_1 \omega_2 +m_1 m_2 +\hat{q}^2)+(m_1\omega_2+ m_2 \omega_1))\\&&
\nonumber a_2= -(m_1\omega_2+ m_2 \omega_1)-(\omega_1 \omega_2+ m_1 m_2 -\hat{q}^2)\\&&
\nonumber a_3= -(\omega_1+\omega_2 +m_1 +m_2)\\&&
\nonumber a_4= (\omega_1-\omega_2) +(m_1 -m_2)\\&&
\nonumber b_1= M (\omega_1\omega_2+m_1 m_2 +\hat{q}^2 -m_1\omega_2- m_2 \omega_1),\\&&
\nonumber b_2= -(m_1\omega_2 +\omega_1 m_2) +(\omega_1\omega_2 +m_1 m_2 -\hat{q}^2),\\&&
\nonumber b_3= (\omega_1 +\omega_2)-(m_1 +m_2);\\&&
b_4= -(\omega_1- \omega_2)+ (m_1- m_2).
\end{eqnarray}

\end{document}